\documentclass[8.5pt,twoside,twocolumn]{article}
\usepackage[super,sort&compress,comma]{natbib}
\usepackage{mhchem}
\usepackage{times,mathptm}
\usepackage{sectsty}
\usepackage{balance}
\usepackage{amsmath}
\usepackage{amssymb}

\usepackage{graphicx} %eps figures can be used instead
\usepackage{lastpage}
\usepackage[format=plain,justification=justified,singlelinecheck=false,font=small,labelfont=bf,labelsep=space]{caption}
\usepackage{fancyhdr}
\begin{document}

\thispagestyle{plain}
\fancypagestyle{plain}{
\fancyhead[L]{\includegraphics[height=8pt]{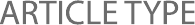}}
\fancyhead[C]{\hspace{-1cm}\includegraphics[height=20pt]{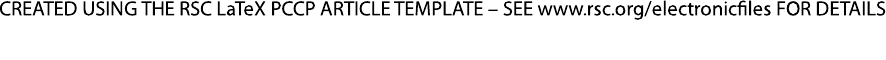}}
\fancyhead[R]{\includegraphics[height=10pt]{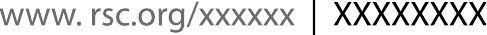}\vspace{-0.2cm}}
\renewcommand{\headrulewidth}{1pt}}
\renewcommand{\thefootnote}{\fnsymbol{footnote}}
\renewcommand\footnoterule{\vspace*{1pt}% 
\hrule width 3.4in height 0.4pt \vspace*{5pt}}
\setcounter{secnumdepth}{5}

\makeatletter
\def\subsubsection{\@startsection{subsubsection}{3}{10pt}{-1.25ex plus -1ex minus -.1ex}{0ex plus 0ex}{\normalsize\bf}}
\def\paragraph{\@startsection{paragraph}{4}{10pt}{-1.25ex plus -1ex minus -.1ex}{0ex plus 0ex}{\normalsize\textit}}
\renewcommand\@biblabel[1]{#1}
\renewcommand\@makefntext[1]% 
{\noindent\makebox[0pt][r]{\@thefnmark\,}#1}
\makeatother
\renewcommand{\figurename}{\small{Fig.}~}

\sectionfont{\large}
\subsectionfont{\normalsize}

\fancyfoot{}
\fancyfoot[LO,RE]{\vspace{-7pt}\includegraphics[height=9pt]{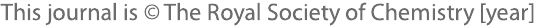}}
\fancyfoot[CO]{\vspace{-7.2pt}\hspace{12.2cm}\includegraphics{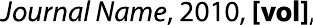}}
\fancyfoot[CE]{\vspace{-7.5pt}\hspace{-13.5cm}\includegraphics{RF.pdf}}
\fancyfoot[RO]{\footnotesize{\sffamily{1--\pageref{LastPage} ~\textbar  \hspace{2pt}\thepage}}}
\fancyfoot[LE]{\footnotesize{\sffamily{\thepage~\textbar\hspace{3.45cm} 1--\pageref{LastPage}}}}
\fancyhead{}
\renewcommand{\headrulewidth}{1pt}
\renewcommand{\footrulewidth}{1pt}
\setlength{\arrayrulewidth}{1pt}
\setlength{\columnsep}{6.5mm}
\setlength\bibsep{1pt}

\twocolumn[
  \begin{@twocolumnfalse}
\noindent\LARGE{\textbf{Procedure to construct a multi-scale
    coarse-grained model of DNA-coated colloids from experimental
    data}}
\vspace{0.6cm}

\noindent\large{\textbf{Bianca M. Mladek$^{\ast}$\textit{$^{a,b}$},
    Julia Fornleitner\textit{$^{c}$}, Francisco
    J. Martinez-Veracoechea\textit{$^{a}$}, Alexandre
    Dawid\textit{$^{d}$}, and Daan
    Frenkel\textit{$^{a}$}}}\vspace{0.5cm}
%Please note that \ast indicates the corresponding author(s) but no footnote text is required. 

\noindent\textit{\small{\textbf{Received Xth XXXXXXXXXX 2010, Accepted Xth XXXXXXXXX 20XX\newline
First published on the web Xth XXXXXXXXXX 200X}}}

\noindent \textbf{\small{DOI: 10.1039/b000000x}}
\vspace{0.6cm}
%Please do not change this text.

\noindent \normalsize{

  We present a quantitative, multi-scale coarse-grained model of DNA
  coated colloids. The parameters of this model are transferable and
  are solely based on experimental data.  As a test case, we focus on
  nano-sized colloids carrying single-stranded DNA strands of length
  comparable to the colloids' size. We show that in this regime, the
  common theoretical approach of assuming pairwise additivity of the
  colloidal pair interactions leads to quantitatively and sometimes
  even qualitatively wrong predictions of the phase behaviour of
  DNA-grafted colloids. Comparing to experimental data, we find that
  our coarse-grained model correctly predicts the equilibrium
  structure and melting temperature of the formed solids. Due to
  limited experimental information on the persistence length of
  single-stranded DNA, some quantitative discrepancies are found in
  the prediction of spatial quantities. With the availability of
  better experimental data, the present approach provides a path for
  the rational design of DNA-functionalised building blocks that can
  self-assemble in complex, three-dimensional structures.}
\vspace{0.5cm}
 \end{@twocolumnfalse}
  ]

\section{Introduction}

\footnotetext{\textit{$^{a}$~Department of Chemistry, University of
    Cambridge, Lensfield Road, Cambridge, CB2 1EW, United
    Kingdom. \\ $^{b}$~Department of Structural and Computational
    Biology, Max F. Perutz Laboratories GmbH, University of Vienna,
    Dr. Bohr-Gasse 9, 1030 Vienna, Austria. Tel: 0043 1 4277 52272 ;
    E-mail: bianca.mladek@univie.ac.at\\ $^{c}$~Institute for Complex
    Systems, Forschungszentrum J\"ulich, 52428 J\"ulich,
    Germany\\ $^{d}$~Universit\'{e} Joseph Fourier Grenoble 1/CNRS,
    Laboratoire Interdisciplinaire de Physique UMR 5588, Grenoble,
    38041, France}}

In the pursuit of designing materials that self-assemble into specific
target structures suitable building blocks have to be found with
interactions that drive the formation of these structures. The
availability of such tailor-made nano-structured materials could open
the way to many interesting applications~\cite{Whi02}. In order to
program self-assembly into nano-sized building blocks, it is crucial
that the interactions between these building blocks can be tuned. One
class of potential `programmable' building blocks are colloidal
particles functionalised with polymers. Such particles can be designed
in many different shapes, ranging in sizes from nm to $\mu$m
\cite{Glo07}. Moreover, the precise choice of their polymeric coating,
i.e.~type, length, flexibility, grafting density and architecture of
the polymers, allows for additional freedom in tuning the interactions
between the particles. Among such systems, DNA-coated colloids
(DNACCs) have received special attention \cite{Gee10,Mir96,Ali96},
mainly because the technology exists to produce specific DNA strands
quickly and cheaply.  These colloidal particles carry short
single-stranded (ss) DNA sequences (``sticky ends'') connected to
inert, grafted polymers (``spacers''). Three-dimensional aggregates of
such colloids can then be formed due to the highly specific and
temperature-reversible hybridisation of complementary sticky ends;
these are either carried by different species of colloids or are part
of so-called linker sequences that bridge between different colloids.

The aggregation behaviour of DNACCs can indeed be influenced via the
properties of the colloids, their polymeric coating, as well as the
solution in which the particles are immersed. Experiments of nano-
\cite{Par08,Nyk08,Xio09,Mac10,May10,Sun11} and micron-sized
\cite{Bia05,Kim06,Cas12} DNACCs have shown that self-assembly of {\it
  simple} spatially ordered structures, such as bcc or fcc crystals,
is possible.  However, applications such as photonic band-gap
materials would require non-close-packed crystals of low coordination,
such as the diamond structure \cite{Ho90}, and despite recent progress
in the field \cite{Cig10,Mac11,Wan12}, the {\it design} of arbitrarily
complex ordered structures is presently still challenging. One crucial
factor is that DNACCs tend to assemble more readily into amorphous
aggregates than into spatially ordered structures \cite{Var12}. The
reason is that the attractions between the DNACCs are strongly
dependent on the external conditions such as ionic strength or
temperature \cite{Jin03,Gee10,Dre10}. Exquisite control over these
parameters is thus needed to help DNACCs to anneal into ordered
structures. Unless we improve our ability to design DNACCs that
assemble readily into the desired target structure, the practical use
of these building blocks remains limited.

It is for this reason that the use of coarse-grained models is
explored both in theoretical approaches
\cite{Tka02,Lic06a,Dre10,Tka11,Var12a} and in computer simulations
\cite{Sta06,Boz08,Dai10,Mar10,Mar11,Var11,Leu11,Sca11,Kno11,
  Chi12,Li12,Mla12}.  These models allow for a fast and efficient
exploration of new design principles of DNACCs. This opens the way to
develop strategies for crystals to form in broader temperature windows
\cite{Mog12a} and offer greater freedom in the design of DNACCs and
the structures they form
\cite{Lic06,Dai10,Tin10,Tka11,Mar11,Kno11,Ang12,Var12a,Cas12}.

Existing models typically range from highly simplified ones
(e.g.~lattice models \cite{Boz08,Tin10} or pair interaction
approaches \cite{Tka02,Sca11}) to sophisticated models featuring
explicit modelling of the DNA hybridisation
\cite{Sta06,Mar10,Mar11,Var11,Leu11,Kno11,Chi12,Li12,Mla12}. In
addition, many models exploit the elastic properties of the DNA
strands: very long strands can be described by scaling laws
\cite{Boz08,Mar10}, while short strands of double-stranded (ds) DNA
can be represented as rigid rods \cite{Dre10,Leu10,Leu11,Mog12a}. Many
of the existing models are qualitative and focus on the generic
features of DNACC self-assembly -- typically, these models do not aim
to describe any specific DNACC system and hence do not exploit the
full available experimental information about the building blocks.

However, for the computer-aided design of DNACCs, quantitative, but
computationally tractable models of DNACCs are needed. By comparison
to the qualitative models mentioned above, models of quantitative
predictive power are rare \cite{Leu10,Rog11,Mog12,Rog12}, and they are
most successful at describing micron-sized colloids covered with short
dsDNA strands, where a description of the systems via pair
interactions determined from simulations has proven
successful. However, the DNACCs that have, thus far, shown most
promise for crystallisation are the ones for which the radius of
gyration of the ssDNA strands, $R_g$, is of comparable size to the
radius of the colloids, $R_C$. In this regime, the modelling
strategies that are successful for larger colloids cannot be applied:
the strands are usually too flexible to be approximated as rigid rods
but too short for polymer scaling laws to apply.

In a recent Letter~\cite{Mla12} we showed that multi-stage coarse
graining can be used to describe the phase behaviour of nano-sized
DNACCs functionalised with ssDNA. The present manuscript describes in
detail the methodology that we have developed to arrive at such a
multi-stage coarse-grained model.  The text is organised as
follows: In the first three sections, we present three steps of
coarse-graining in which we identify the key degrees of freedom that
determine the phase behaviour of DNACCs: we develop our most
detailed model of DNACCs based on experimental data in
Sec.~\ref{sec:lvl1}. Based on simulations of this model, we derive the
``core-blob model'', the second level of coarse-graining, in
Sec.~\ref{sec:lvl2}. This model, in turn, allows us to perform the final
step of coarse-graining and calculate effective interactions
(Sec.~\ref{sec:lvl3}). The expected reliability of the effective
interactions to predict the phase behaviour of DNACCs is assessed in
Sec.~\ref{sec:3body}. Finally, we calculate the phase diagram of the
chosen DNACCs within both the core-blob model and the effective
interaction approach in Sec.~\ref{sec:xtals}. In the Appendices, we
detail technical aspects of the present work.

\section{Stage 1: Model with explicit DNA chains} \label{sec:lvl1}
\subsection{General outline}
To develop the most detailed level of description of DNACCs, a
suitable model for the ssDNA strands tethered to the colloid's surface
has to be chosen. ssDNA is not a simple polymer: it is prone to form
hairpin structures and knots.  An accurate description of such
substructures could be achieved by fully atomistic simulations, which
are computationally feasible at most for small systems of DNACCs
covered with few, short DNA strands
\cite{Lee09a,Ngo12}. Alternatively, rather detailed, coarse grained
models of DNA such as developed in Ref. \citenum{Oul11} could be
employed. But while this model makes the detailed study of
hybridisation between several strands of DNA feasible \cite{Oul10}, it
is still computationally too time-consuming to be employed for a
system of hundreds or thousands of colloids, each covered with dozens
of ssDNAs.  Fortunately, the formation of ssDNA loops and knots is
expected to play a minor role for DNACCs where commonly DNAs are
chosen that are not self-complementary; we thus neglect this effect
and model the ssDNA strands as freely jointed, charged chains. This
model captures the most important contribution to the behaviour of the
ssDNA strands which stems from the electrostatic repulsion of the
DNA's sugar-phosphate backbone~\cite{Zha01}. In view of the high
Young's modulus of ssDNA \cite{Zha01}, the segments of every freely
jointed, charged chain are chosen to have a fixed Kuhn length $l_{\rm
  Kuhn} = 2p_{\rm ss}$, where $p_{\rm ss}$ is the persistence length
of ssDNA.  The number of Kuhn segments $n_K$ per chain is determined
as
\begin{equation}
n_K = \left \lfloor \frac{l_{\rm contour}}{l_{\rm Kuhn}} \right \rfloor.
\end{equation}
Here, $l_{\rm contour} = (N_b-1) b_0$ is the contour length of the
ssDNA strand, with $N_b$ the number of bases per strand, and $b_0$ the
interbase distance in ssDNA. The symbol $\lfloor \dots \rfloor$
denotes the floor function. We stress that since $l_{\rm Kuhn} > b_0$,
each Kuhn segment represents several nucleotides. Consequently, our
model cannot capture the precise base sequence of the ssDNA strands;
sequence dependent effects, such as base stacking, are captured only
in an averaged way in the choice of $b_0$ (see below).

The conformation of a freely jointed, charged chain is defined by the
positions $\{ {\bf r}_{0},...,{\bf r}_{n_K}\}$ of the $n_K+1$ vertices
of the chain.  We approximate the continuous charge of the backbone by
effective charges sitting at each of these vertices; two vertices $i$
and $j$ at distance $r_{ij} = |{\bf r}_{i}-{\bf r}_{j}|$ interact with
each other via a Debye-H{\"u}ckel interaction, $\Phi_{ij}^{\rm ele}$,
given by
\begin{equation}\label{dh}
\Phi_{ij}^{\rm ele} = \frac{q^2}{4 \pi \varepsilon_0 D}
\frac{e^{-\kappa r_{ij}}}{r_{ij}},
\end{equation}
where $D$ is the dielectric constant of the solvent and
$\varepsilon_0$ is the vacuum permittivity (in SI units). The charge
per vertex, $q$, is approximated as
\begin{equation}
q = \frac{l_{\rm contour}}{(n_K+1)} \nu,
\end{equation}
with the effective line charge density of ssDNA, $\nu$. Finally, the
inverse Debye screening length, $\kappa$, in Eqn.~\ref{dh} is given by
\begin{equation}
\kappa = \sqrt{\frac{2 \beta I N_A e^2 10^3}{D \varepsilon_0}}\;,
\end{equation}
where $N_A$ is Avogadro's number and $e$ is the elementary
charge. $\beta = 1/k_{\rm B}T$, where $k_{\rm B}$ denotes Boltzmann's
constant and $T$ stands for the temperature (all in SI units). The
ionic strength $I$ of the solution in which the DNACCs are immersed is
given in mol/l and the factor $10^3$ stems from converting mol/l to SI
units.

\begin{figure}%[!b]
\includegraphics[width=8.3cm]{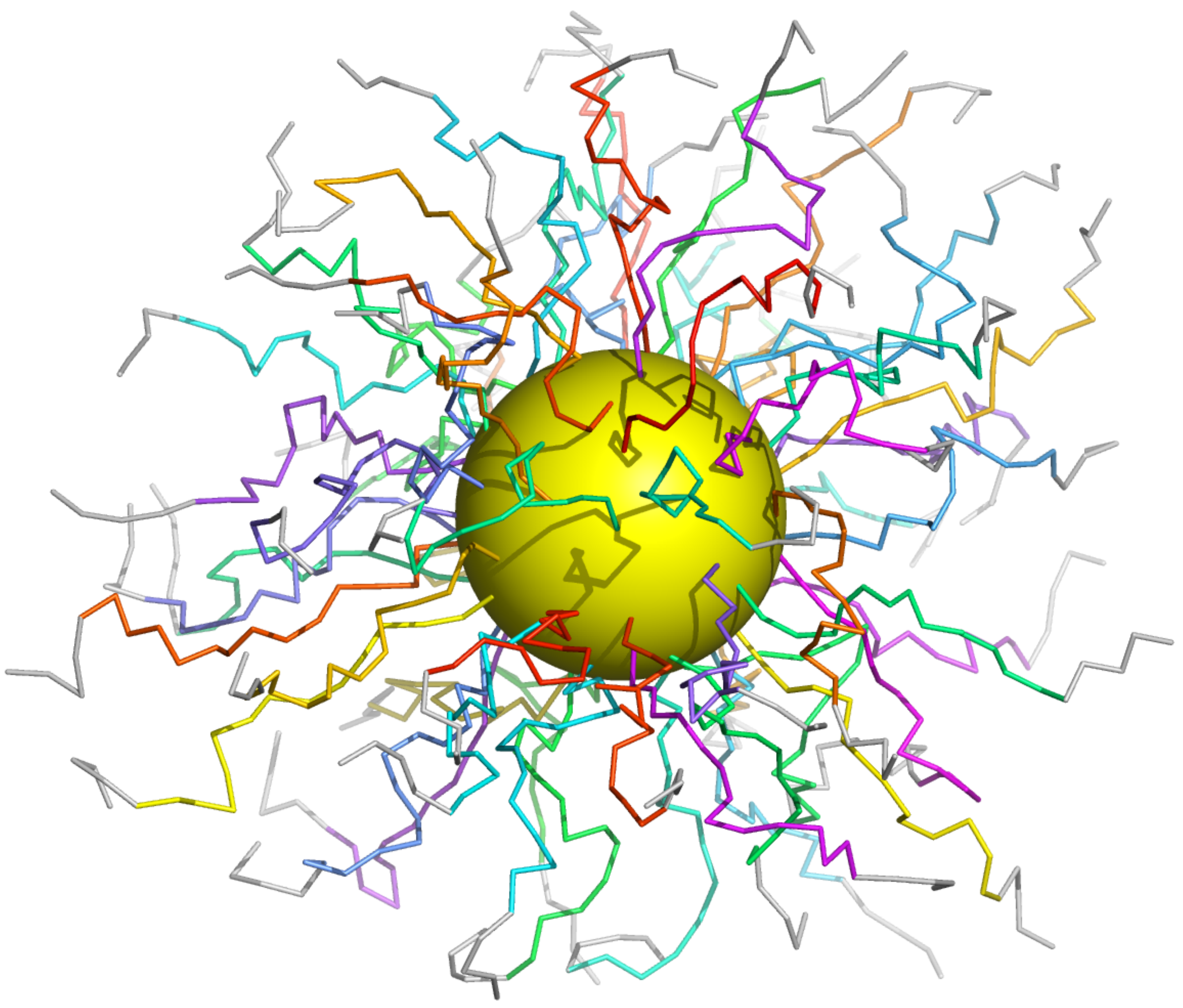}
\caption{(Colour online) Simulation snapshot of our most detailed
  model of DNACCs: a hard sphere colloid (yellow) of $R_C=6$ nm is
  dressed with 60 strands of freely jointed, charged chains of 18 Kuhn
  segments (corresponding to 65 bases; various colours) at the solvent
  conditions given in Ref. \citenum{Nyk08} and at $T=25^{\circ}$C. The 4
  last segments model the sticky ends and are plotted in grey.}
\label{fig:level1}
\end{figure}

The $n_K$ segments of each ssDNA strand can be divided into two
classes: the number of sticky end segments $n_{\rm se}=\lfloor
(N_{b,{\rm se}}-1)b_0/l_{\rm Kuhn} \rfloor$, calculated from the
number of sticky end bases $N_{b,{\rm se}}$; and $n_{\rm sp} =
n_K-n_{\rm se}$, the number of spacer segments. A total of $N_{\rm
  str}$ ssDNA strands are then grafted to the colloid: we attach the
DNA strands at their first vertex ${\bf r}_0$, neglecting in our model
the hexane-thiol group by which the ssDNA strands are experimentally
tethered and which is estimated to have a end-to-end length of $\sim
0.8$ nm \cite{Ngo12}. The colloid, in turn, is modelled as a hard
sphere of radius $R_C$ that cannot be penetrated by the vertices of
the DNA strands (see Fig.~\ref{fig:level1}). In experiments
\cite{Nyk08, Hur06}, colloids are typically maximally loaded with DNAs
and therefore we assume that the anchoring points are uniformly
distributed on the surface of the colloid and---in accordance with
experimental evidence \cite{Dymtro}---that they cannot diffuse.

\subsection{Chosen values}
Knowing the experimental conditions under which the reference
experiments were performed, we can determine the values of all
variables introduced in the last section. As a proof-of-concept of our
method, we choose to study system V from Ref. \citenum{Nyk08}, where a
symmetric, binary mixture of gold nano-colloids (labelled $A$ and $B$)
of radius $R_C\sim 6$ nm were studied. All colloids were coated with
$\sim$ 60 ssDNA strands of $N_b=65$ bases, out of which $N_{b,{\rm se}}
= 15$ bases constitute the sticky end. $A$ and $B$ colloids only
differed in their sticky end sequences, which were complementary to
allow for direct hybridisation between the unlike colloids. These
DNACCs were assembled in a solution of 0.01 mol/l phosphate buffer,
0.2 mol/l NaCl at pH = 7.1 \cite{Nyk08}. Using the
Henderson-Hasselbach equation \cite{Hen08}, the concentration $c_i$ of all ion
species $i=1,\dots,n$ in the solution can be calculated; then, the
ionic strength, $I$, is given as, $I = \frac{1}{2} \sum
\limits_{i=1}^n c_i z_i^2$, where $z_i$ denotes the charge number of ion
species $i$. For the present system \cite{Nyk08}, we find that
$I=0.21$ mol/l.

Next, we need to set the persistence length of ssDNA. While the
persistence length of dsDNA is well known, its value for ssDNA is less
well established: a broad range of experimentally measured values has
been published, varying between 0.75 nm and almost 10 nm
\cite{Kuz01}. Here, we use a value of 0.75 nm, in accordance with the
studies on which our ssDNA model is based \cite{Zha01}. According to
Ref. \citenum{Tin97}, $p_{\rm ss}$ depends on the ionic strength and the
dependence is approximately described by $p_{\rm ss}[\AA] \sim 4 I{\rm
  [mol/l]}^{-1/2}$. Therefore, at the ionic strength of our reference
experiment (i.e.~$I = 0.21$mol/l), this results in $p_{\rm ss} \sim
0.87$ nm, which is reasonably close to the value of 0.75 nm used
here. In a similar way, values reported for the interbase distance
$b_0$ vary considerably, since they depend on the precise DNA sequence
under study and the physical conditions of the solution in which the
DNA is immersed. While an inter-phosphorus distance of $0.59$ nm has
been established for ssDNA \cite{Smi96}, stacking of bases leads to an
interbase distance that is on average shorter~\cite{MogXX,Mil99};
motivated by the findings of Ref. \citenum{Tin97}, we choose a value of
$b_0=0.43$ nm. Therefore, the contour length of the ssDNA strands used
in the present study is $l_{\rm contour} \sim 27.5$ nm. With both
$l_{\rm contour}$ and $l_{\rm Kuhn} = 2p_{\rm ss} = 1.5$ nm ready at
hand, we find that $n_K=18$, $n_{\rm se}=4$, and consequently $n_{\rm
  sp}=14$. For an isolated DNA strand, we find a radius of gyration of
$R_g \sim 0.7 R_C$, thus the radius of gyration of the DNA strands is
indeed of the order of the colloidal size.

The effective line charge density $\nu$ is interpolated from Table 1
in Ref. \citenum{Zha01} and we find $\nu(I=0.21$ mol/l; ssDNA$) =
2.07$ $e$/nm, translating to a charge of $q\sim 3$ $e$ per vertex. In our
study, we use $D=80$ for all temperatures and the Debye screening
length varies from 0.67 nm (at 25 $^\circ$C) to 0.71 nm (at 65 $^\circ$C).

\subsection{Simulations}
To study the behaviour of an isolated DNACC, we implement Monte Carlo
simulations utilising crankshaft and pivot moves to equilibrate the
ssDNA chains.  Every few Monte Carlo sweeps, we also try to regrow
whole chains by employing configurational bias Monte Carlo simulations
\cite{Fre02}.

\begin{figure}%[!b]
\includegraphics[width=8.3cm]{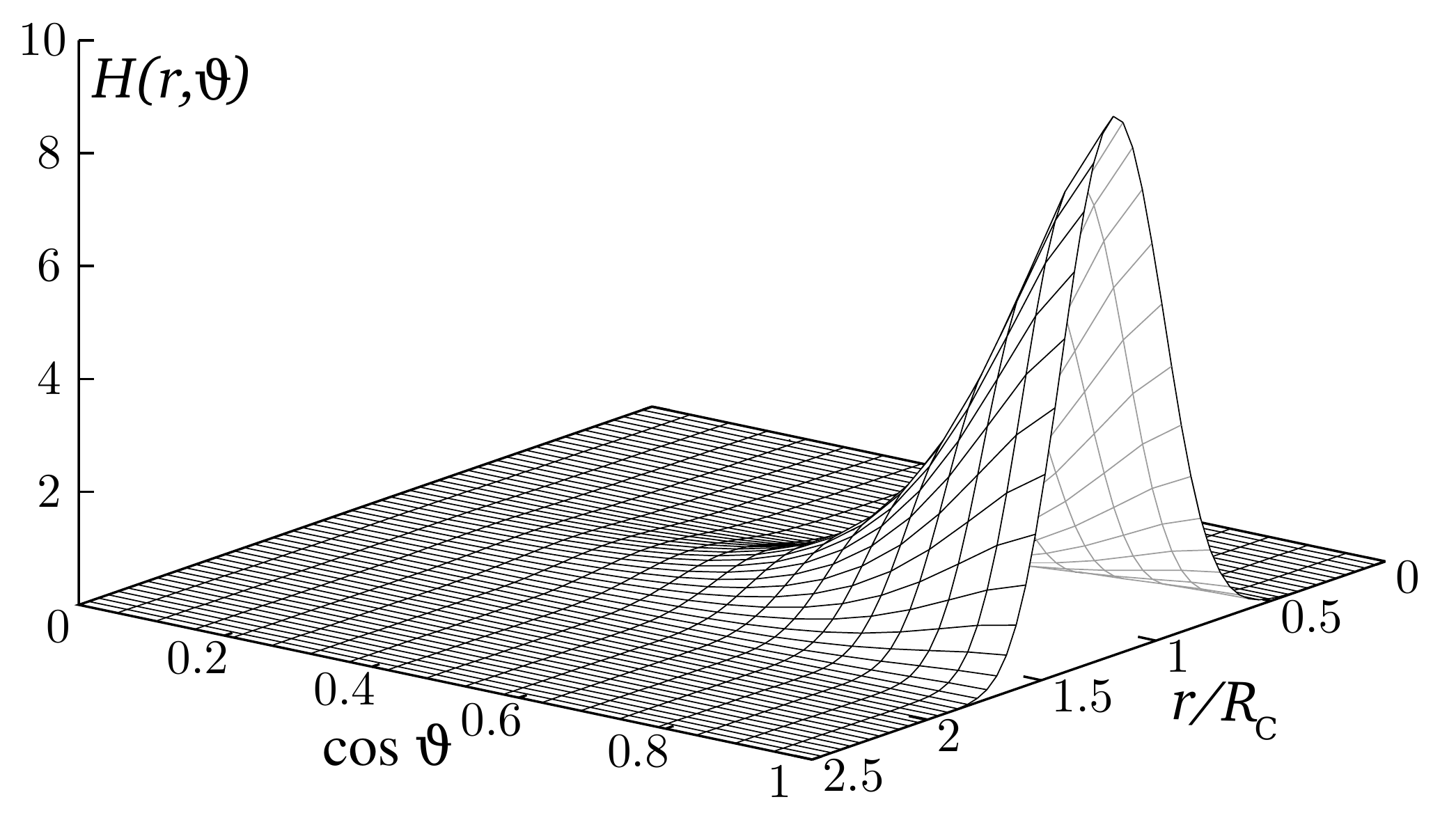}
\caption{The height distribution $H(r,\vartheta)$ of reactive ends, as
  function of (a) the distance $r$ of the centre of mass of the last
  $2n_{\rm se}$ Kuhn segments of a strand, ${\bf r}_{\rm end}$, from
  its anchoring point on the colloid, ${\bf r}_{\rm anchor}$; and (b)
  the deviation in angle $\vartheta$ between the vectors ${\bf r}_{\rm
    end}-{\bf r}_{\rm anchor}$ and the connection vector from the
  colloid's centre to the anchoring point. This distribution is peaked
  at ($r \sim 1.2 R_C$, $\vartheta = 0$).}
\label{fig:histo}
\end{figure}

These simulations then allow us to gain insight into the height
distribution $H(r,\vartheta)$ of the ends of the DNA strands. For
convenience, we measure this distribution as a function of two
parameters: (a) the distance $r$ of the centre of mass of the last
$2n_{\rm se}$ Kuhn segments of a strand, ${\bf r}_{\rm end}$, from its
anchoring point on the colloid, ${\bf r}_{\rm anchor}$. The choice of
$2n_{\rm se}$ will be motivated in Sec.~\ref{sec:lvl2}; (b) the
deviation in angle $\vartheta$ between the vectors ${\bf r}_{\rm
  end}-{\bf r}_{\rm anchor}$ and the connection vector from the
colloid's centre to the anchoring point. This distribution captures
how sticky ends are restricted in their movement due to the fact that
the ssDNA strands are tethered to the colloid and due to neighbouring
DNA strands, showing a peak at ($r \sim 1.2 R_C$, $\vartheta = 0$)
(Fig.~\ref{fig:histo}).

\begin{figure}%[!b]
\includegraphics[width=8.3cm]{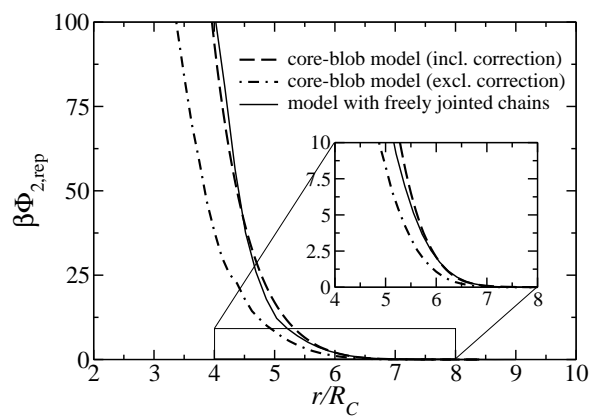}
\caption{The steric repulsion between two DNACCs, $\beta \Phi_{\rm 2,rep}$, as
  function of the distance between them and as obtained by the model
  with explicit DNA chains (solid line) and the core-blob model (without
  correction: dash-dotted line, with correction: dashed line). The
  repulsion is found to be independent of temperature. The inset shows
  a close-up of the functions for distances $r/R_C \gtrsim 5$.}
\label{fig:cb_fix}
\end{figure}

In addition, we use a modification of Widom's particle insertion
(mWPI) technique \cite{Mla10} to determine the steric repulsion
$\Phi_{\rm 2, rep}(r,T)$ between two DNACCs separated by distance $r$
and at temperature $T$ in the zero density limit. We find $\beta
\Phi_{\rm 2,rep}$ to be temperature-independent over a wide range of
temperatures from T=25 $^\circ$C to 75 $^\circ$C
(Fig.~\ref{fig:cb_fix}, solid line).

The large number of degrees of freedom with which DNACCs are described
in the present model would render large-scale simulations of crystals
of DNACCs unfeasible. Also, the present model would call for a binding
scheme of the sticky ends, where several Kuhn segments align to form
the dsDNA stretch~\cite{Sta06,Var11,Kno11,Li12,Chi12}, which is
computationally rather expensive. We therefore refrain from
implementing binding between complementary ssDNA sequences in the
present model and rather develop a more coarse-grained model---which
we term core-blob model---in the next section.

\section{Stage 2: Core-blob model} \label{sec:lvl2}
\begin{figure}%[!b]
\includegraphics[width=8.3cm]{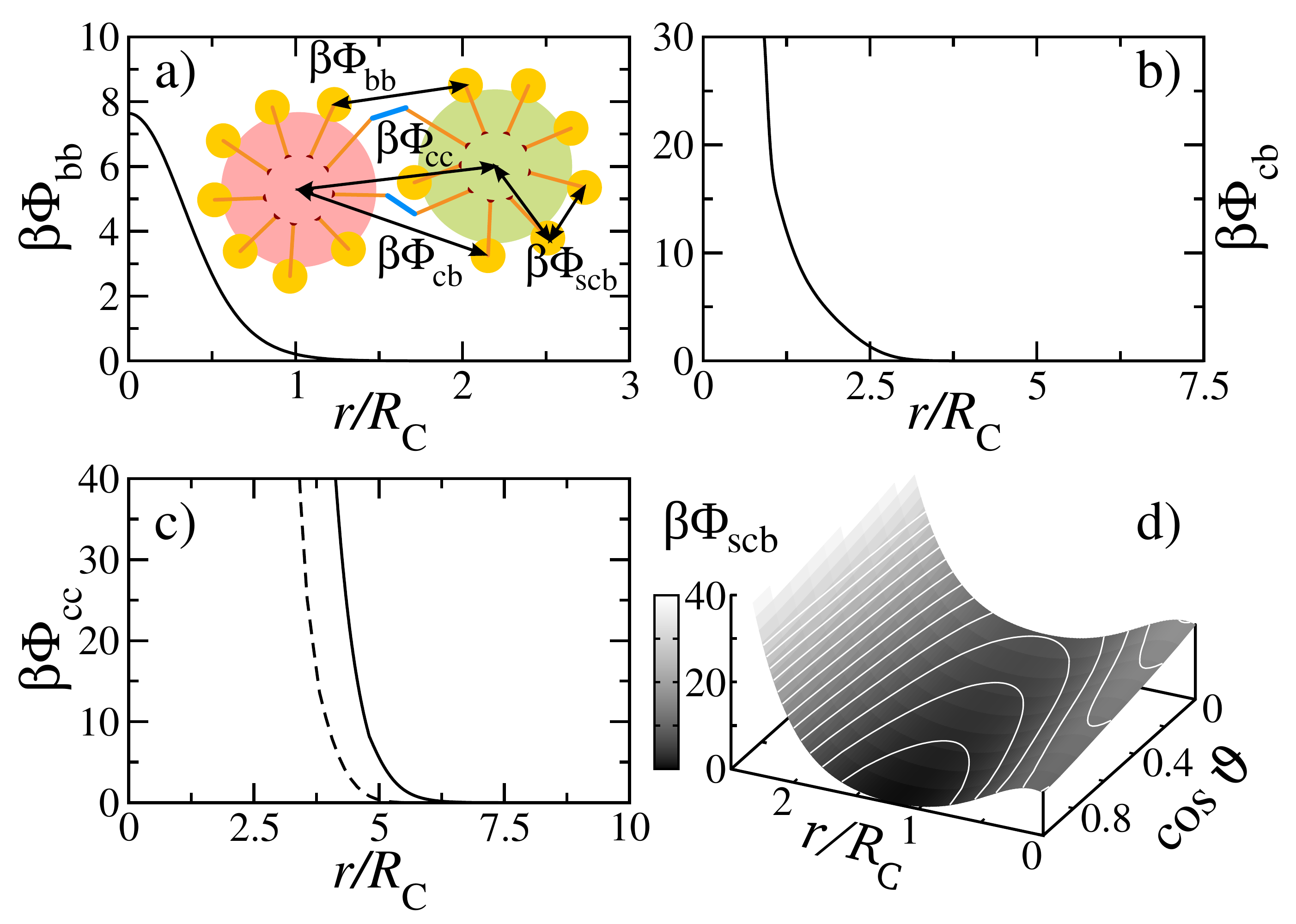}
\caption{(Colour online) The core-blob model, in which sticky ends are
  modelled as blobs and the spacers and colloids constitute a core
  [see inset in a)].  a) Blobs on different colloids interact via
  $\beta \Phi_{\rm bb}$; b) a blob with the core of another colloid
  via $\beta \Phi_{\rm cb}$; c) different cores via $\beta \Phi_{\rm
    cc}$ (dashed line: uncorrected potential; solid line: corrected
  potential); and d) blobs with their own colloid via $\beta \Phi_{\rm
    scb}$ (Isolines shown every 2.5 $k_BT$). Further, blobs on
  colloids of different identity can bind and form short stretches of
  dsDNA [sketched as blue rods in the inset in a)].}
\label{fig:model}
\end{figure}

The aim of the core-blob model is to arrive at the simplest possible
model of DNACCs which preserves (i) the steric repulsion $\Phi_{2,\rm
  rep}$ between two isolated DNACCs and (ii) the height profile of
sticky ends with respect to their colloid as obtained from the model
of explicit DNA chains (see Sec.~\ref{sec:lvl1}). We therefore model each
of the $N_{\rm str}$ sticky ends as an entity called ``blob'', which
is grafted to the surface of the colloid at fixed anchoring points. In
this, the last $2n_{\rm se}$ segments of every freely jointed, charged
chain constitute a blob; then---to a first approximation---the blob's
centre represents the connection point between the sticky end and its
spacer.  The gold colloid and the remaining $n_K-2n_{\rm se}$ segments
of all $N_{\rm str}$ strands form the ``core'', leading to a model of
$N_{\rm str}+1$ separate entities. The model is defined by four
different interactions (Fig.~\ref{fig:model}) which we derive from the
model of explicit DNA chains via Monte Carlo simulations using
mWPI~\cite{Mla10} and biased simulations~\cite{Fre02}:

\begin{enumerate}
\item[(i)] the repulsive interaction $\Phi_{\rm bb}(r)$ acting between
  two blobs tethered to {\it different} colloids and a distance $r$
  apart is approximated as the interaction of two isolated
  (=non-tethered) freely jointed, charged chains of length $2n_{\rm
    se}$~\cite{Bol01a}. As anticipated from studies of polymers, this
  potential is of Gaussian shape (see e.g.~\cite{Bol01a}). However,
  due to the finite length and the charge carried by the ssDNA, the
  repulsion found here is considerably stronger than the 2$k_BT$
  characteristic of polymers in the scaling regime (see
  Fig.~\ref{fig:model}a);

\item[(ii)] the interaction $\Phi_{\rm cb}(r)$ of a blob with the core
  of {\it another} colloid separated by distance $r$. We approximate
  this potential by simulating a single freely jointed, charged chain
  of $2n_{\rm se}$ segments interacting with a bare core, i.e.~a hard
  sphere grafted with chains of length $n_K-2n_{\rm se}$ (see
  Fig.~\ref{fig:model}b);

\item[(iii)] the interaction $\Phi_{\rm cc}(r)$ between two cores at
  distance $r$ is estimated as the zero density repulsion between two
  colloids each grafted with chains of length $n_K-2n_{\rm
    se}$ (see Fig.~\ref{fig:model}c);

\item[(iv)] the interaction of a single sticky end with {\it all} the
  remains of its {\it own} colloid (i.e.~the core and all other blobs)
  cannot trivially be split into a repulsive and a tethering
  contribution due to intricate multi-body contributions.  In an
  isolated DNACC, we therefore determine the full multi-body potential
  $\Phi_{\rm scb}(r,\vartheta)$ from the height distribution of sticky
  ends in the underlying model as $\Phi_{\rm scb}(r,\vartheta) = -\log
  H(r,\vartheta)$ (see Fig.~\ref{fig:model}d). By construction, this
  potential guarantees the preservation of the height profile of
  sticky ends with respect to the model of explicit DNA chains in the
  regime of dilute solutions of DNACCs (see Sec.~\ref{sec:lvl1}).
\end{enumerate}

To evaluate the reliability of the core-blob model, we determine the
steric repulsion $\Phi_{2,\rm rep}(r)$ between two isolated DNACCs
within this model via mWPI~\cite{Mla10} and compare the results to our
findings from the model of explicit DNA chains. We find that the
core-blob model underestimates $\Phi_{2,\rm rep}(r)$ (see
Fig.~\ref{fig:cb_fix}, dash-dotted line). The reason for this
discrepancy can be traced back `to the fact that the core-core
repulsion $\Phi_{\rm cc}$ is too soft (Fig.~\ref{fig:model}c, dashed
line), since this potential should also include a multi-body
contribution from the sticky ends tethered to the spacer chains, which
cannot be captured by $\Phi_{\rm bb}$ alone. We therefore introduce a
correction to $\Phi_{\rm cc}$ (Fig.~\ref{fig:model}c, solid line) by
simulating the repulsion between two colloids dressed with chains of
length $n_K-n_{\rm se}$ instead of $n_K-2n_{\rm se}$. Then, the
core-blob model recovers $\Phi_{2, \rm rep}$ with sufficient accuracy
for all distances. Especially, we find good agreement for distances
$r> 5.5 R_C$ (see Fig.~\ref{fig:cb_fix}, dashed line) which includes
the range of experimentally observed next neighbour distances, $r = a
\sqrt{3}/2 \gtrsim 6.1 R_C$, where the measured CsCl lattice spacings
$a \gtrsim 42.5$ nm~\cite{Nyk08}. Defining the colloidal packing
fraction $\eta = \frac{4\pi}{3} R_C^3 N/V$ (with $V$ the volume of the
conventional unit cell and $N$ the number of DNACCs in this unit cell)
and assuming the experimentally observed CsCl structure to be the
thermodynamically stable structure, we can determine the packing
fraction up to which our model shows high reliability as $\eta =
\frac{4\pi}{3} R_C^3 2/a^3$. Using $a = 2\dot 5.5 /\sqrt{3} R_C$, we
find $\eta \lesssim 0.033$.

Finally, the hybridisation of complimentary sticky ends has to be
modelled via a suitable Monte Carlo move. In this, we have to account
for binding of initially unbound sticky ends, breakage of initially
bound sticky ends, as well as for the change of binding partner for an
already hybridised sticky end. Further, we wish to use the
experimentally measured data on the DNA hybridisation free energy.
Since the persistence length of dsDNA far exceeds that of ssDNA, we
model the hybridised sticky ends as a (volumeless) rigid rod of fixed
length $\mathcal{L}= n_{\rm se}l_{\rm Kuhn}$. For simplicity, we
ignore the change in inter-base distance between ssDNA (0.43 nm) and
dsDNA (0.34 nm). Binding is possible between a chosen (bound or
unbound) blob $i$ on a given colloid and all unbound blobs $j$
tethered to unlike colloids and within a distance of approach $r_{ij}
< \mathcal{L}$. Upon binding, the reaction partner $j$ is moved to a
distance $\mathcal{L}$ from $i$ along the connection line ${\bf
  r}_{ij}$. Since bound blobs cannot move independently anymore,
binding leads to the loss of a degree of freedom which is reintroduced
upon unbinding by placing $j$ along ${\bf r}_{ij}$ with a probability
of $r_{ij}^2$, thereby guaranteeing detailed balance.  The
probabilities for each possible bound state ${ij}$ and the unbound
state $i$ to occur are determined by their respective weights $W$:
\begin{equation}
W_{ij} = \frac{K}{\frac{4 \pi}{3} {\mathcal L}^3 \rho_0} \exp(-\beta
U_{ij}),\;,
\end{equation}
and
\begin{equation}
W_i = \exp(-\beta U_i),\;,
\end{equation}
where $\rho_0$ is the standard density of 1 mol/l, $ U_{ij}$ is the
potential energy of the state where $i$ is bound to $j$ and $U_i$ the
potential energy of the state where $i$ is unbound; $K$ is the
equilibrium binding constant of two sticky ends and is connected to
the hybridisation free energy $G_{\rm hyb}$ via $K=\exp\left(-\beta
G_{\rm hyb}\right)$; $G_{\rm hyb}$ (and thereby $K$) depends on the
nucleotide sequence of the sticky ends and it is temperature- and
salt-dependent; it can be approximated as the hybridisation free
energy of two sticky ends free in solution (e.g.~via
\texttt{DINAMelt}~\cite{Mar05}). The values for $G_{\rm hyb}$ used
here are given in Tab.~\ref{tab}. The present binding move is
justified in more detail in the Appendix A.

\begin{center}
\begin{table}
  \begin{tabular}{| c | c || c | c |} 
\hline
$T$ [$^{\circ}$C] & $G_{\rm hyb}$ [$k_{\rm B}T$]  &  $T$ [$^{\circ}$C] & $G_{\rm hyb}$ [$k_{\rm B}T$] \\  \hline
50  & -14.12 & 61.4 & -9.03\\ \hline
55 & -11.83 & 62.1 & -8.71 \\ \hline
55.8 & -11.47 & 63.2 & -8.23 \\ \hline
56.9 & -10.98 & 64.3 & -7.75\\ \hline
58.0 & -10.49 & 65.1 & -7.44\\ \hline
59.1 & -10.00 & 66.2 &-6.97\\ \hline
60.2 & -9.51 & 66.9 &-6.67\\ \hline
  \end{tabular}
\caption{The values of the hybridisation free energy $G_{\rm hyb}$
  [$k_{\rm B}T$] of the DNA strands used in system V in
  Ref. \citenum{Nyk08} at different temperatures (according to
  \texttt{DINAMelt}~\cite{Mar05}): the DNA spacer sequence is given as
  5'-TTT TTT TTT TTT TTT TTT TTT TTT TTT CGT TGG CTG GAT AGC TGT GTT
  CT-3'. Sticky ends on $A$-colloids read 5'-TAA CCT AAC CTT CAT-3',
  while on $B$-colloids the complementary sequence is found, 5'-ATG
  AAG GTT AGG TTA-3'. Values were determined at a ionic strength of
  0.21 mol/l.}
\label{tab}
\end{table}
\end{center}

\section{Stage 3: Effective interactions}\label{sec:lvl3}
We use the core-blob model to calculate the pair interactions between
two DNACCs in the zero density limit. Interactions between like
colloids (i.e.~$AA$ and $BB$) are described by the purely repulsive,
temperature-independent steric repulsion between two DNACCs, $\beta
\Phi_{\rm 2, rep}$, calculated above. The interaction of unlike
colloids (i.e.~$AB$) additionally features an attractive potential
$\Phi_{\rm 2, hyb}$ stemming from hybridisation of DNA strands. Thus,
\begin{equation}
\beta \Phi_2^{AA/BB}(r) =  \beta \Phi_{\rm 2, rep} (r)
\end{equation}
and
\begin{equation}
\beta \Phi_2^{AB}(r,T) = \beta \Phi_{\rm 2, rep} (r) + \beta \Phi_{\rm 2, hyb} (r,T).
\end{equation}
$\beta \Phi_{\rm 2, hyb}$ can be obtained by evaluating
\begin{equation}\label {eqn:ti}
  \beta \Phi_{2,{\rm hyb}}(r,T) = -\int_{\beta G_{\rm hyb}}^\infty {\rm
    d}(\beta G'_{\rm hyb}) \left<\zeta(r)\right>_{K=\exp(-\beta G'_{\rm hyb})},
\end{equation}
as described in Ref. \citenum{Leu11}. Here, $\zeta(r)$ is the number
of DNA bridges formed between the two colloids at fixed distance $r$,
and $\left< \dots \right>$ denotes the statistical average. In
practise, the integration is performed between $G_{\rm hyb}$ of
interest and a value sufficiently large for sticky ends not to
hybridise anymore.

Due to the temperature-dependence of $G_{\rm hyb}$, the depth of the
minimum of $\Phi_2^{AB}(r,T)$ varies strongly with temperature.  For
instance, a change in temperature from 62.1 to 56.9 $^\circ$C results
in a drop in the minimum of $\Phi_2^{AB}$ of roughly 20 $k_{\rm B}T$
(corresponding to 13 kcal/mol). This strong temperature dependence of
the DNA-mediated attraction explains the difficulty in crystallising
the DNACCs: only in a narrow temperature range around $\sim65 ^\circ$C
is the minimum in $\Phi_2^{AB}(r,T)$ shallow enough to allow for the
formation {\it and} breakage of DNA links.  Upon lowering $T$, the
bonds that form cannot break anymore and the system gets stuck in
disordered aggregates, even if an ordered structure is
thermodynamically stable \cite{Whi02}.

To assess the predictive power of the core-blob model, it would be
desirable to compare $\Phi_2^{AA/BB}$ and $\Phi_2^{AB}$ to
experimental results. But while such potentials can be experimentally
determined for {\it micron}-sized colloids by using optical tweezers
\cite{Bia05,Rog11}, the same is not feasible for nano-colloids.  Other
experimental validation techniques will be required. Thus, validation
of our model against experimental data will only be studied later, by
comparing the computed and experimentally determined phase diagrams.

\begin{figure}%[!b]
\includegraphics[width=8.3cm]{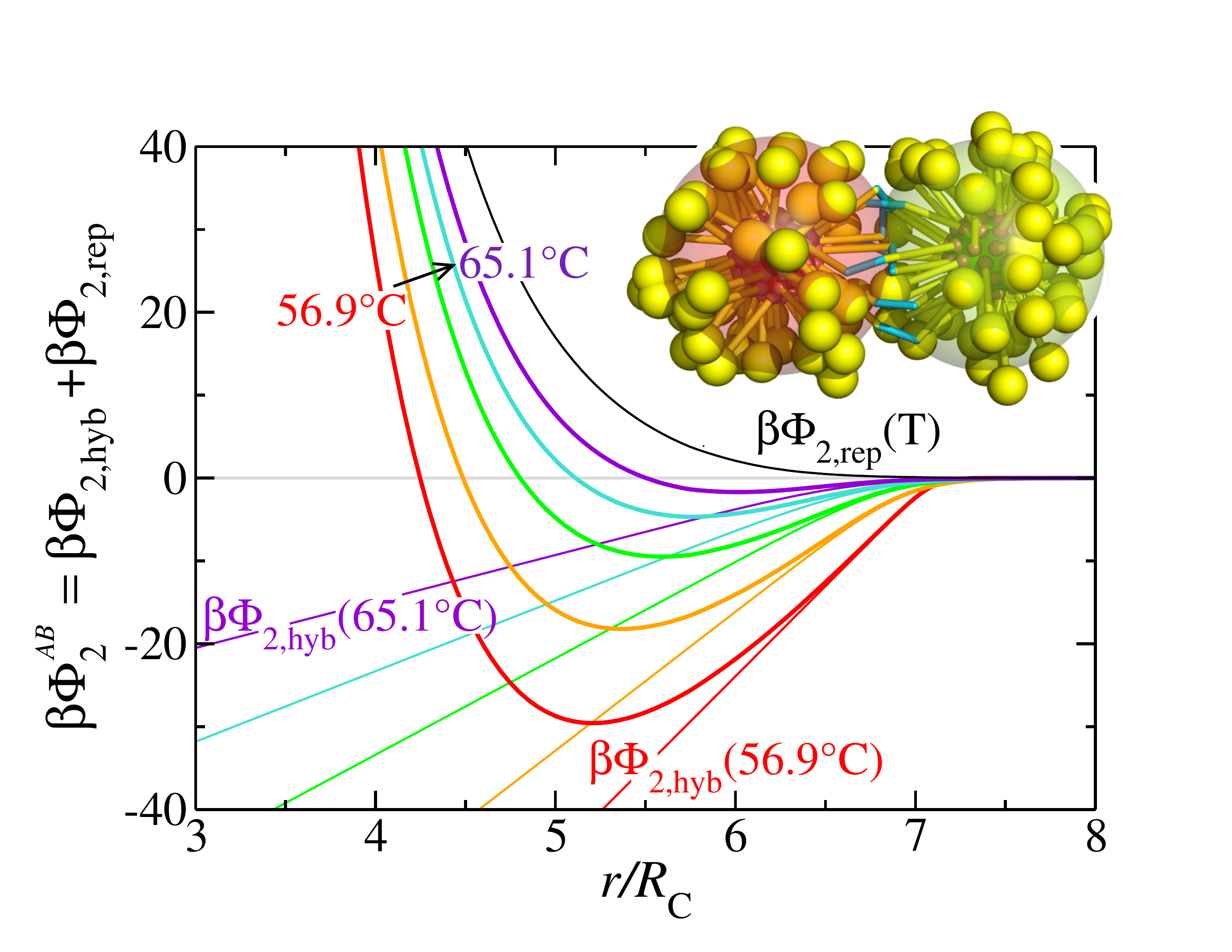}
\caption{(Colour online) The effective interaction between unlike
  colloids $\beta \Phi_{\rm 2, rep}^{AB}$ (bold lines) is the sum of
  the temperature-independent steric repulsion $\beta \Phi_{\rm 2,
    rep} (= \beta \Phi_{2}^{AA/BB})$ between two colloids, and a
  temperature-dependent attractive potential $\beta \Phi_{\rm 2, hyb}$
  arising from hybridisation of DNA strands. Data are shown for
  different temperatures (top to bottom: 65.1$^{\circ}$C,
  63.2$^{\circ}$C, 61.4$^{\circ}$C, 59.1$^{\circ}$C, and
  56.9$^{\circ}$C). The inset shows a simulation snapshot of an $A$
  colloid (green) interacting with a $B$ colloid (red). Unhybridised
  sticky ends are shown as yellow spheres, while DNA bridges are shown
  as blue rods. The translucent spheres indicate the average position
  of the sticky ends.}
\label{fig:phi2}
\end{figure}

 Thus far, our approach has allowed us to compute the effective pair
 potential between DNACCs.  Pairwise additive interactions are
 typically used to model DNACCs as structureless particles in
 theoretical studies (e.g.~Ref. \citenum{Tka02}) and also in some
 computational studies (e.g.~Ref. \citenum{Sca11}). However, in the
 regime where $R_g \sim R_C$ we may expect that the assumption of
 pairwise additivity breaks down.  With the present model we can
 quantify the importance of such many-body interactions.

\section{Three-body interactions}\label{sec:3body}
Using the core-blob model we test if the three-body interactions of a
system of two $A$ colloids and one $B$ colloid can be written as the
sum of the various two-body contributions. It can be anticipated that
differences between the three-body interaction and the sum of the
two-body contributions will mainly arise from a competition of the two
$A$ colloids for the sticky ends of the $B$ colloid and will therefore
crucially depend on the arrangement of the DNACCs with respect to each
other. A linear arrangement of the colloids, with the $B$ colloid
positioned between the two $A$ colloids, is expected to lead to little
discrepancy since in the regime studied here ($R_g \sim R_C$), the DNA
strands are too short to reach complementary sticky ends at the back
of the DNACC they face.

However, in the typical arrangements of DNACCs that occur in a
crystalline environment, many-body effects are more likely to arise.
In a crystal, a given colloid is typically surrounded by several
DNACCs of the other species that all compete for the strands of the
central colloid. It is therefore interesting to study the three-body
interaction $\Phi_3^{ABA}=[\Phi_3^{BAB}]$ for three DNACCs arranged in
an equilateral triangle of side-length $r$ and test if the following
relation holds
\begin{equation}
\beta \Phi_{3}^{ABA} (r,T) \stackrel{?}{=} 3 \beta \Phi_{2,{\rm
    rep}}(r) + 2 \beta \Phi_{2,{\rm hyb}}(r,T).
\end{equation}

\subsection{Methods}
In analogy to the $AB$ two-body interactions, also the $ABA/BAB$
three-body interaction can be split into a contribution $\Phi_{\rm 3,
  rep}$ stemming from steric repulsions between the DNACCs, and an
attractive part $\Phi_{\rm 3, hyb}$ arising from DNA strand
hybridisation, i.e.
\begin{equation}
\beta \Phi_3^{ABA}(r,T) = \beta \Phi_{\rm 3, rep} (r,T) + \beta
\Phi_{\rm 3, hyb} (r,T).
\end{equation}

 We first calculate $\Phi_{\rm 3, rep}$ by generalising the
 mWPI~\cite{Mla10}: we place three {\it non-reactive} DNACCs,
 i.e. DNACCs with non-binding sticky ends, in an equilateral triangle
 of side-length $r = r_{\rm max}$ bigger than the expected range of
 the interactions. Then we repeatedly reduce the side-length to
 $r-\Delta r$ and measure $\left<\exp\left[-\beta \Delta U(r
   \rightarrow r-\Delta r)\right]\right>$, where $\Delta U(r
 \rightarrow r-\Delta r)$ is the change in potential energy between
 the three DNACCs due to the move. In this way, we move the DNACCs
 together. Then, $\beta\Phi_{3, \rm rep}(r) = - \sum_{r'=r_{\rm
     max}}^{r+\Delta r} \log \left< \exp \left[ -\beta \Delta U(r'
   \rightarrow r'-\Delta r)\right]\right>$.

To determine $\Phi_{3,{\rm hyb}} (r)$, we arrange the three DNACCs in
an equilateral triangle of fixed side-length $r$. Measuring the total
number of DNA bridges formed at this distance, $\zeta(r)$, we can
calculate $\Phi_{3,{\rm hyb}} (r)$ as
\begin{equation}
\beta \Phi_{3,{\rm hyb}}(r) = -\int_{\beta G_{\rm hyb}}^\infty {\rm
  d}\left(\beta G'_{\rm hyb}\right)
\left<\zeta(r)\right>_{K=\exp(-\beta G'_{\rm hyb})},
\end{equation}
in analogy to the determination of $\Phi_{2,\rm hyb}$ (see
Eqn.~\ref{eqn:ti}).
\subsection{Results}
\begin{figure}%[!b]
\includegraphics[width=8.3cm]{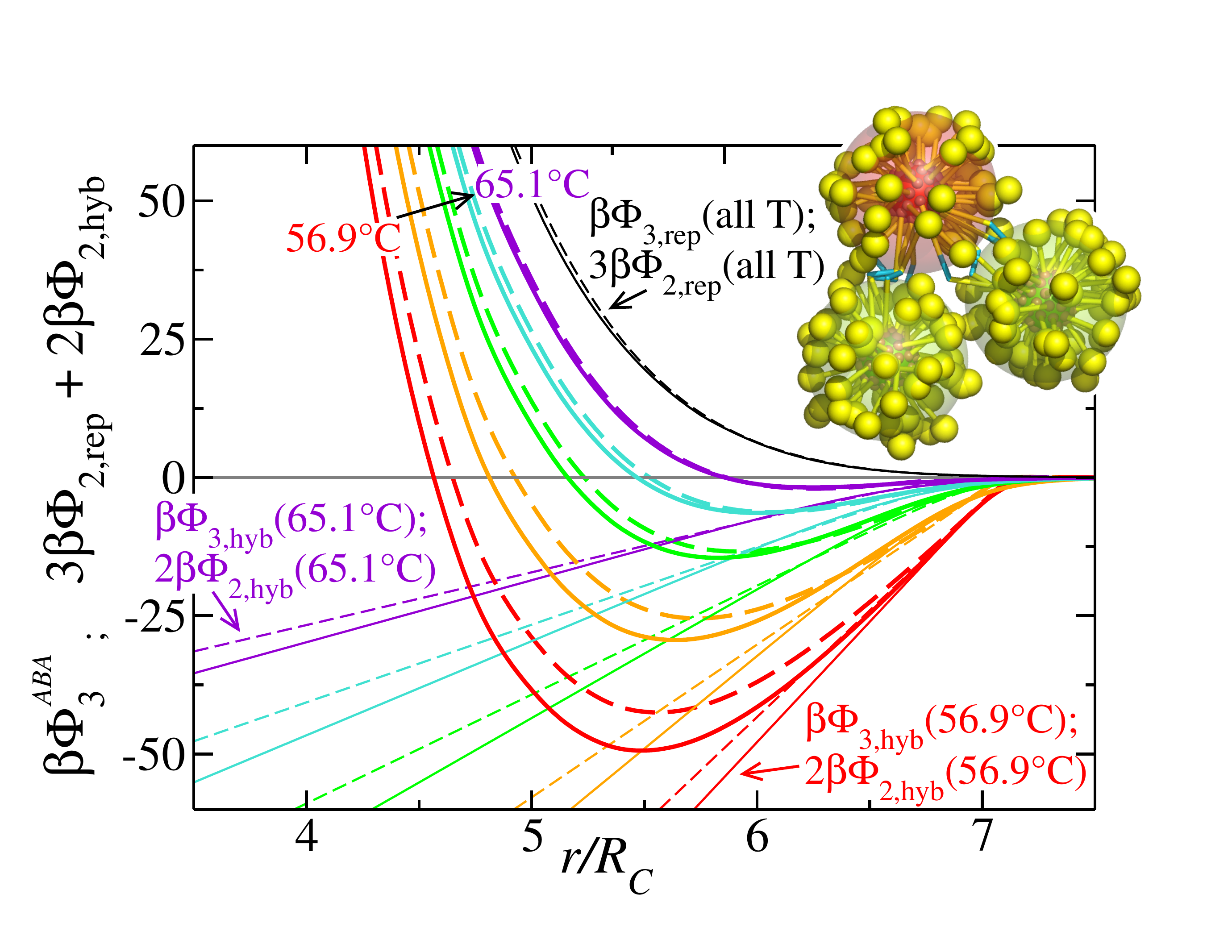}
\caption{(Colour online) The three-body effective interaction $\beta
  \Phi_3(r)$ (bold dashed lines) of two $A$ and one $B$ colloids
  arranged in an equilateral triangle compared with the sum of the
  two-body contributions $3\beta \Phi_{2, {\rm rep}}+2 \beta \Phi_{2,
    {\rm hyb}}$ (bold solid lines) at 65.19$^{\circ}$C,
  63.29$^{\circ}$C, 61.49$^{\circ}$C, 59.1$^{\circ}$C, and
  56.9$^{\circ}$C (top to bottom). Further, the repulsive contribution
  and the attractive hybridisation contributions to the effective
  interactions are shown explicitly in thinner lines (two-body
  contributions: solid lines, three-body contributions: dashed
  lines). The inset shows a simulation snapshot. $A$ colloids are
  shown as green spheres, and the $B$ colloid as a red
  sphere. Unhybridised sticky ends are shown as yellow spheres, while
  DNA bridges are sketched as blue rods. The translucent spheres
  indicate the average position of the sticky ends.}
\label{fig:lvl3}
\end{figure}

As can be seen from Fig.~\ref{fig:lvl3}, the repulsive three-body
potential $\beta \Phi_{\rm 3, rep} (r)$ is, to a good approximation,
equal to the sum of the two-body contributions $3 \beta \Phi_{\rm 2,
  rep}(r)$. However, pairwise additivity does not hold for the
attractive part of the three-body potential, $\beta \Phi_{\rm 3,
  hyb}$.  As anticipated, the two $A$ colloids increasingly compete
for the available sticky ends of $B$ as the colloids are moved closer
together and fewer bonds can form for each of the two $AB$ pairs than
in an isolated, single $AB$ pair. This overestimation of formed DNA
bridges within the pair potential approach directly translates to an
overestimate of the depth of the attraction between the colloids
(cf. Eqn.~\ref{eqn:ti} and see Fig.~\ref{fig:lvl3}) and consequently
an underestimate of the position of the minimum in the attraction. We
therefore expect that an analysis of DNACC crystals within the pair
potential framework will predict more compact crystals than found
experimentally \cite{Nyk08}.

\section{Phase behaviour of DNA coated colloids} \label{sec:xtals}
To assess the predictive power of both the core-blob model and the
pair potential approach, we study the phase behaviour of DNACCs by
implementing free-energy calculations within both approaches. The
results are then compared to data available from experiments
\cite{Nyk08}, such as the stable crystal structure, its lattice
constant and the melting temperature of these crystals.

\subsection{Crystal structure prediction}
As was already known to Ostwald \cite{San84}, observing spontaneous
formation of a crystal does not imply that the observed structure has
the lowest free energy.  Rather, we have to consider the thermodynamic
stability of all possible crystal structures. To identify credible
candidates for the most stable crystal structure, we use optimisation
techniques based on genetic algorithms~\cite{Hol75, Got05}. The
structures that the genetic algorithm identifies as plausible are then
considered in the free-energy calculations.  We adapt a search
strategy for 2D binary mixtures~\cite{For08,For09} to 3D, augmenting
it with a parametrisation of search space that excludes {\it a priori}
configurations with overlapping colloids~\cite{Pau09}. The lattice
parameters describing the crystal structures are encoded in binary
individuals and a random crossover is employed as mating scheme.
Mutations take place with a rate of $0.05$. We limit our search to
symmetric $AB$-mixtures and lattice structures with up to eight
particles per unit cell. Particles interact via their respective pair
interactions (see Sec.~\ref{sec:lvl2}), which we first fit to
analytical functions (see Appendix B). Calculations are run at
constant pressure $P$, so that the volume fraction $\eta$ enters the
optimisation as an independent parameter.  Structures are optimised
with respect to the Gibbs free energy of the system $G=U+PV-TS$, with
$U$ the internal energy, $V$ the volume of the system, and $S$ the
entropy. In the genetic algorithm, entropy is neglected, and hence the
Gibbs free energy is equal to the enthalpy of the
structures. Temperature-dependence of the system only enters our
approach via the temperature-dependence of the pair potential; this
limitation in treating entropic effects necessitates the subsequent
free energy calculations (see Sec.~\ref{fecalc}). To determine the
minimum enthalpy configuration at a given pressure we evaluate 1000
generations of a population of 50 individual ordered structures
each. Details on the general working principles of the method can be
found in Ref. \citenum{Got05}.

The genetic algorithm calculations predict the CsCl (B2) structure as
the most stable one for low colloidal packing fractions $\eta$, while
it also predicts a competing NaTl (B32) structure for higher
$\eta$. The latter structure is of general interest since it is
composed of two interpenetrating diamond structures. If it were
possible to remove one of the two colloidal species in a post-assembly
modification step, a diamond structure could be created
\cite{Ho90}. Such assembly strategies are explored by
e.g.~substituting the gold colloids in one of the two species of
DNACCs by organic compounds \cite{Cig10}.  However, in the present
system an experimental distinction between CsCl and NaTl structures
would prove challenging: here, $A$ and $B$ colloids only differ by
their sticky ends while X-ray scattering only detects the gold
colloids. As a result, both CsCl and NaTl structures would
experimentally be detected as bcc arrangements.

\begin{figure}%[!b]
\includegraphics[width=8.3cm]{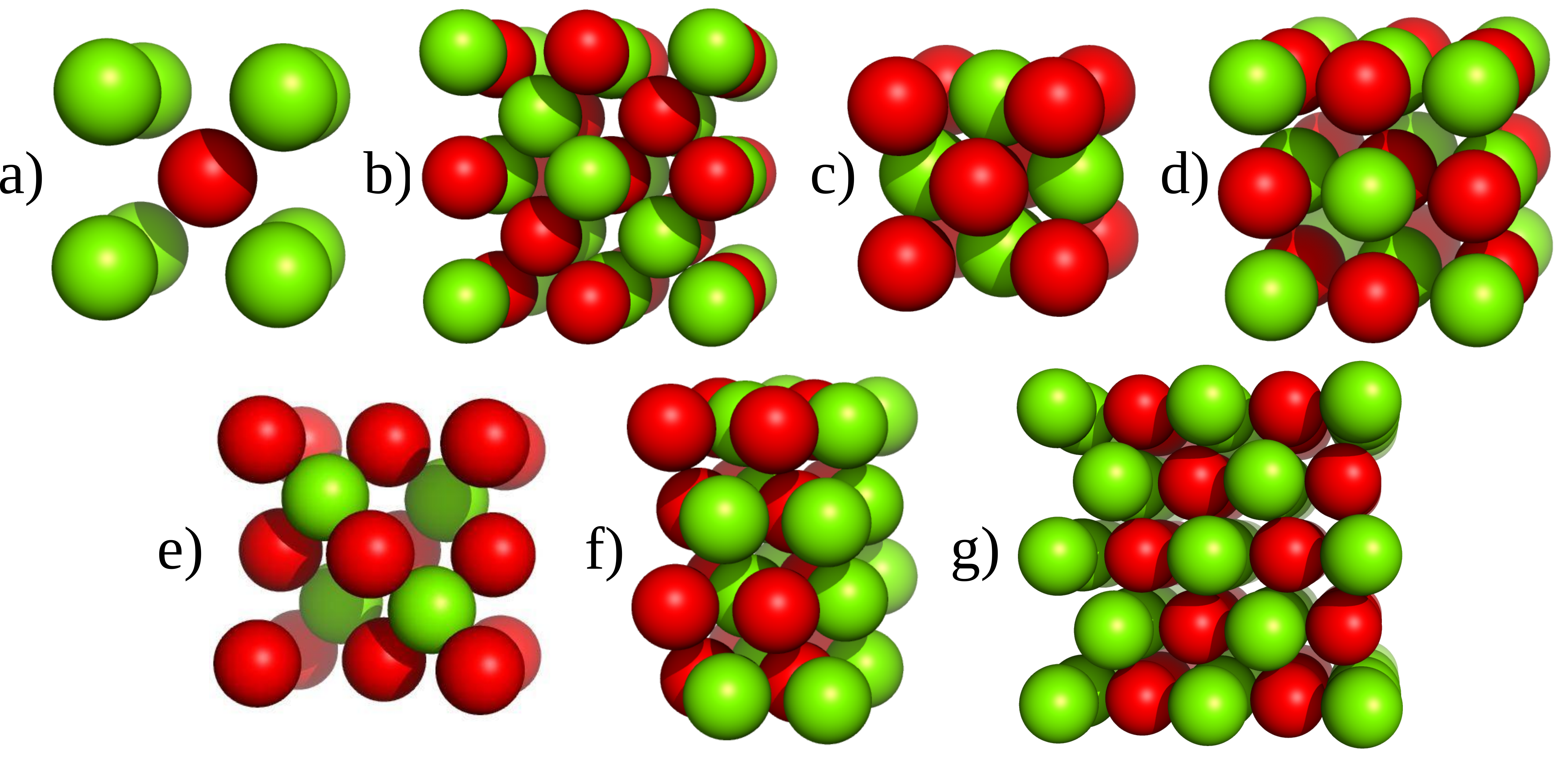}
\caption{(Colour online) The different crystal structures considered
  in this study. a) CsCl; b) NaTl; c) CuAu; d) NaCl; e) ZnS; f)
  'straight' hcp; and g) AuCd. Different colloidal species are
  coloured red and green; for clarity, the DNA strands have been
  omitted.}
\label{fig:xtals}
\end{figure}

Apart from the CsCl and NaTl structures, we chose to consider a few
more candidate structures: CuAu (L$1_0$), NaCl (B1), `straight' hcp
(s-hcp) \cite{Sca11}, ZnS (B3; diamond) \cite{Tka02}, AuCd (B19), as
well as substitutionally disordered CsCl and CuAu crystals
\cite{Par08,Kno11} (Fig.~\ref{fig:xtals}).

\subsection{Free energy simulations}\label{fecalc}
To study the phase behaviour of DNACCs and to determine the stable
crystal structure, we have to determine the free energies $F$ of all
candidate structures mentioned above. Within the effective pair
potential framework, we calculate $F$ via thermodynamic integration in
the canonical ensemble as detailed in~\cite{Mla07,Mla08} and using
systems of at least $N=1000$ DNACCs. Within the core-blob model, the
thermodynamic integration is achieved in two steps similar to the
approach followed in Ref. \citenum{Mar11}, using systems of more than
$N=100$ DNACCs. In the first step of the integration we arrange $N$
DNACCs in the desired crystal structure at fixed colloidal volume
fraction $\eta$ and temperature $T$ (and thereby fixed binding free
energy $G_{\rm hyb}$; see Tab.~\ref{tab}). We then transform this
DNACC crystal into a system of non-binding DNACCs artificially fixed
to their lattice sites. We achieve this by gradually increasing
$G_{\rm hyb}$ so that sticky ends do not bind anymore; at the same
time, we gradually confine the centres of the colloids to individual,
small cells of volume $v$ around the perfect lattice sites ${\bf
  R}^{\rm LS}$ of the chosen crystal structure by raising a potential
barrier. For a colloid $i$ centred at ${\bf R}_i^C$ and at integration
point $\lambda \in [0,1]$, the potential energy $\phi_{\rm
  barr}^\lambda$ due to the barrier is given as
\begin{equation}
\beta \phi_{\rm barr}^\lambda({\bf R}_i^C) = \begin{cases}
0 & \text{if ${\bf R}_i^C \in \cup_j v({\bf R}_j^{\rm LS})$}\\
\lambda \beta U_{\rm barr} & \text{else}\;,
\end{cases}
\end{equation}
where $\cup_j v({\bf R}_j^{\rm LS})$ is the union of the confining
volumes around all perfect lattice sites ${\bf R}^{\rm LS}$ and
$U_{\rm barr}$ is the maximal height of the barrier. $U_{\rm barr}$ is
chosen sufficiently high for the crystals not to melt during the
thermodynamic integration. The total potential energy of the system
due to the barrier is given as $\beta \Phi_{\rm barr}^\lambda =
\sum_{i=1}^N \beta \phi_{\rm barr}^\lambda({\bf R}_i^C)$. Via
Gauss-Lobatto quadrature \cite{Abr64}, we numerically evaluate
\begin{equation}\label{eqn:f1}
\beta \Delta F_1 = \int \limits_1^0 {\rm d}\lambda \left<\beta
  \Phi_{\rm barr}^\lambda\right> - \int \limits_{\beta G_{\rm
    hyb}}^\infty {\rm d}(\beta G')_{\rm hyb}
\left<\zeta\right>_{K=\exp(-\beta G'_{\rm hyb})},
\end{equation}
where $\Delta F_1$ is the difference in free energy between the
crystal of interest and the crystal of inert, confined DNACCs.  In
practise, the upper limit of the second integral is replaced by a
hybridisation free energy sufficiently large to guarantee that no
hybridisation of the ssNDA sticky ends takes place.  $\zeta$ denotes
the total number of DNA bridges formed in the system.

In the second integration step, we use the lattice-coupling expansion
method \cite{Mei90}: we linearly expand both the crystal of inert
DNACCs and the potential barrier---which guarantees to hold the
crystal in place---i.e.
\begin{equation}
{\bf R}_i^X \rightarrow \gamma {\bf R}_i^X
\end{equation}
with $X=C, {\rm LS}$ and the expansion factor $\gamma \in [1,\infty]$.
In practise, the infinite expansion is approximated by expanding the
system sufficiently for particles not to interact anymore. The free
energy difference between the unexpanded and the expanded crystal of
inert DNACCs is given by
\begin{equation}\label{eqn:f2}
\beta \Delta F_2 = \beta \int\limits_1^{\infty} {\rm d}\gamma \left< {\mathcal{W}}
\right>_\gamma\;,
\end{equation}
where ${\mathcal{W}}$ is a modified virial given as
\begin{equation}
{\mathcal{W}} = \sum \limits_{i<j} {\bf f}_{ij, \gamma} {\bf R}_{ij,\eta}^C.
\end{equation}
Here, ${\bf f}_{ij, \gamma}$ is the force acting between DNACC $i$ and
DNACC $j$ at expansion factor $\gamma$, and ${\bf R}_{ij,\eta}^C ={\bf
  R}_{j,\eta}^C - {\bf R}_{i,\eta}^C $ is the separation of the two
DNACCs at the {\it original} colloidal packing fraction $\eta$
(i.e.~at $\gamma = 1)$~\cite{Mei90}.

 The total free energy per colloid is then given as
\begin{equation}
 \beta F/N = \beta f_{\rm id} + \beta \Delta F_1/N+ \beta \Delta F_2/N,
\end{equation}
where $\beta f_{\rm id}$ is the free energy of an isolated DNACC;
being the same for all crystal structures, this contribution can be
neglected in the determination of the thermodynamically stable crystal
structure.

\subsection{Results}
For all candidate crystal structures, we determine the free energy for
a range of packing fractions, concentrating on the regime where $ T
\lesssim 65$ $^\circ$C. This is the temperature where the effective
interaction $\Phi_2^{}$ develops a minimum indicating that at these
temperatures DNA bridges form between DNACCs. In this regime, crystals
of low packing fraction are expected to be stabilised by DNA
hybridisation.

\begin{figure}%[!b]
\includegraphics[width=8.3cm]{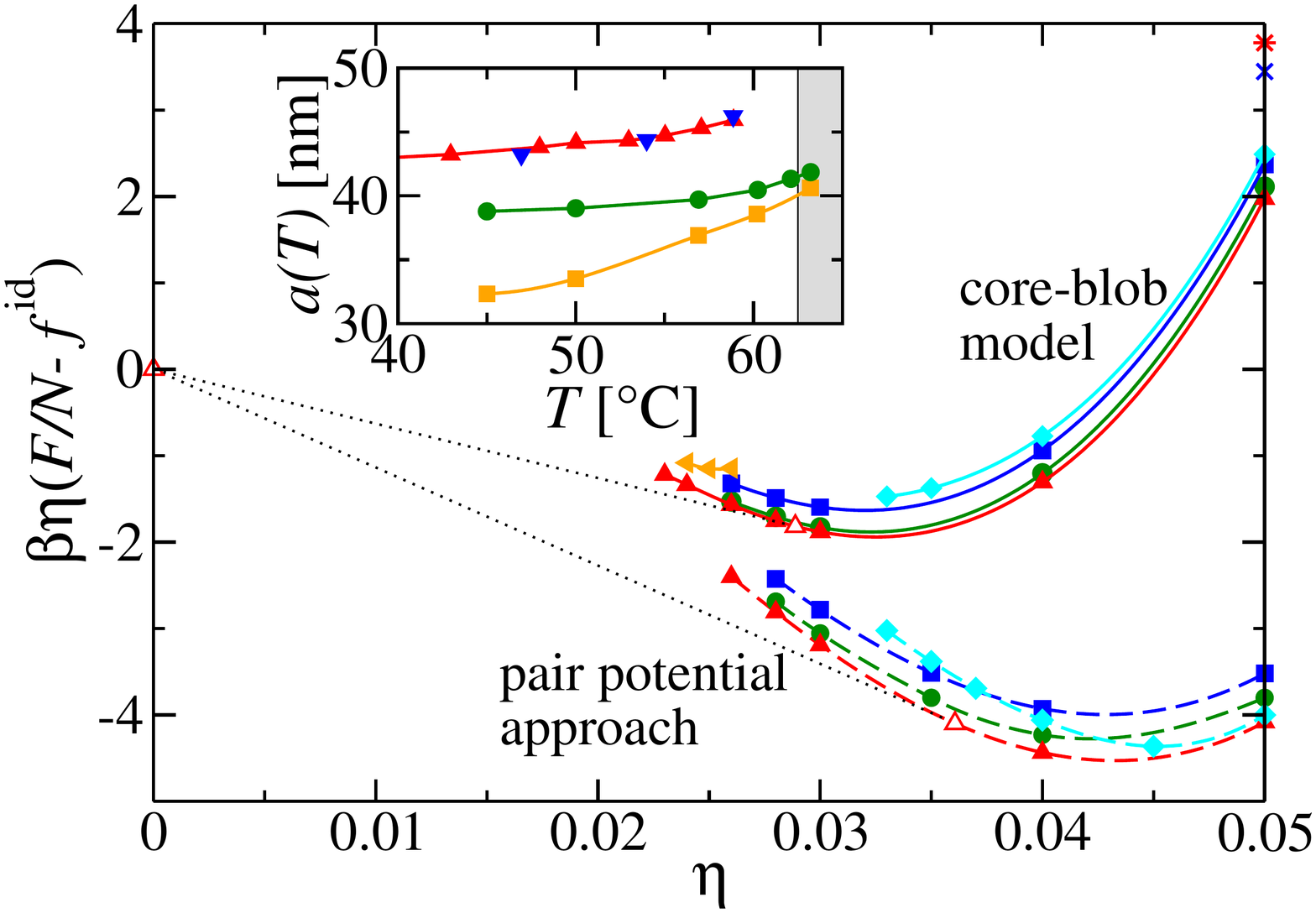}
\caption{(Colour online) The excess free energy $\beta \eta (F/N -
  f^{\rm id})$ as function of the colloidal volume fraction $\eta$ at
  $T=56.9^{\circ}$C. Both the core-blob model (solid lines) and the
  pair potential approach (dashed lines) find the CsCl structure
  ($\blacktriangle$) to be the most stable one. This crystal is in
  equilibrium with a dilute vapour ($\triangle$). The common tangents
  are shown as dotted lines. In contrast, the liquid
  ($\blacktriangleleft$) is found to be only metastable. In close
  competition with the CsCl structure, a range of metastable crystal
  structures is found: s-hcp: $\bullet$, CuAu: $\blacksquare$, NaTl:
  $\blacklozenge$, disordered CsCl: $\ast$, disordered CuAu:
  $\times$. Inset: A comparison of the lattice constant $a$ (in nm) as
  function of temperature $T$ (in $^{\circ}$C) as measured in the
  experiments (heating: $\blacktriangle$, cooling:
  $\blacktriangledown$~\cite{Nyk08}) and as obtained from simulations
  with the core-blob model ($\bullet$) and using the pair potential
  approach ($\blacksquare$). The shaded region indicates the
  temperatures at which crystals are not stable in experiments.}
\label{fig:free_energy}
\end{figure}

We first present results obtained via the core-blob model
(Fig.~\ref{fig:free_energy}, solid lines), initially concentrating on
a temperature of $T=56.9$ $^\circ$C. We find that at all packing
fractions $\eta$ considered, the CsCl structure is the most stable, in
agreement with experimental findings \cite{Nyk08}. In close
competition with the CsCl structure, we find a whole range of
metastable crystal structures, namely s-hcp, CuAu and NaTl. These
structures are mechanically stable for packing fractions of $\eta
\gtrsim 0.026-0.035$; by contrast, the CsCl structure is already found
stable for $\eta \gtrsim 0.023$. At high $\eta \sim 0.065$, a
metastable AuCd phase appears. By contrast, both the NaCl and ZnS
structures are found to be mechanically unstable and melt at all
volume fractions considered. Extrapolation of the collected data
suggests that the NaTl structure would become the stable structure for
$\eta \gtrsim 0.07$. The reason for this transition from CsCl to NaTl
can be understood from studying the mechanism stabilising the latter
structure:
we found that NaTl crystals are mechanically stable only at packing
fractions $\eta \gtrsim 0.035$, where colloids not only form DNA
bridges with next neighbour colloids, but also with the unlike
colloids found in the second coordination shell. By contrast, CsCl
cannot form DNA bridges the the next nearest neighbours since they are
all like colloids. At sufficently high packing fractions beyond $\eta
\sim 0.05$, the excluded volume effects of the DNA strands render
binding to the nearest neighbours incresingly challenging for both
CsCl and NaTl. While this severely limits the amount of possible DNA
bridges for CsCl, NaTl can instead bind to the next nearest
neighbours, allowing it to eventually become the thermodynamically
stable structure around $\eta \gtrsim 0.07$. However, such high volume
fractions cannot be achieved in experiments where crystals form from a
dilute vapour.  Moreover, we stress that the extrapolation should be
taken with a grain of salt as we do not expect our model to be fully
valid at these high densities (see Sec.~\ref{sec:lvl2}).

Next, we study substitutionally disordered crystals. At low volume
fractions $\eta < 0.05$, such crystals have only limited mechanical
stability and few substitutional changes can be sustained by the
crystals. A single substitutional defect, in which only a single pair
of neighbouring $A$ and $B$ colloids exchange sites in an otherwise
perfect crystal increases the free energy of the crystal by $\Delta F
= 0.5k_{\rm B}T$ ($\eta = 0.024$) to 1.6$k_{\rm B}T$ ($\eta = 0.05$)
for CsCl and by $\Delta F = 0.1k_{\rm B}T$ ($\eta = 0.026$) to
1.0$k_{\rm B}T$ ($\eta = 0.05$) for CuAu. At higher $\eta \sim 0.05$,
more substitutional disorder can be stabilised: however, the free
energy of the substitutionally disordered structures is higher than
that of the perfectly ordered crystals.  This can be understood by
considering the contributions to the free energy.  While $\beta \Delta
F_2$ (Eqn.~\ref{eqn:f2}) and the first integral of $\beta \Delta F_1$
(Eqn.~\ref{eqn:f1}) are independent of the distribution of the $A$ and
$B$ colloids over lattice sites, the second integral determining
$\beta \Delta F_1$ (Eqn.~\ref{eqn:f1}) depends crucially on the
average number of DNA bridges $\left<\zeta\right>$ formed in the
crystal. As $\left<\zeta\right>$ is lower in a substitutionally
disordered crystal than in a perfectly ordered crystal, $\beta \Delta
F_1$ is lower for the latter systems. Still, substitutionally
disordered crystals have been seen to form spontaneously in
simulations~\cite{Kno11}. The present results suggest that these
structures are kinetically arrested for $T\lesssim 65$ $^\circ$C.

At high $\eta > 0.1$, it is to be anticipated that the phase behaviour
of DNACCs will increasingly be dominated by the steric repulsion from
the compressed DNA strands as well as excluded volume interactions of
colloids. Then, close-packed structures such as CuAu or s-hcp are
expected to be thermodynamically stable. Since crystals are not
stabilised by DNA hybridisation anymore in this regime,
substitutionally disordered crystals should be favoured over
ordered structures. We emphasise that exploration of this regime is
beyond the scope of the current contribution.

Compared to the results of the core-blob model, the pair potential
approach---as expected---underestimates the free energy by
overestimating the number of DNA bridges formed in the system
(Fig.~\ref{fig:free_energy}, dashed lines). Still, it offers a good
estimate of the range of mechanical stability of the various crystal
structures and predicts the same phase order as the core-blob model,
i.e.~CsCl as most stable structure, followed by metastable s-hcp, CuAu
and NaTl. However, within the pair potential approach, the NaTl
structure is found to out-compete the CsCl structure already around
lower packing fractions of $\eta \gtrsim 0.055$. Structures found to
be mechanically unstable in the core-blob approach (ZnS, CuAu) are
also found unstable within the current framework. Therefore, the pair
potential approach offers an excellent tool for assessing the
mechanical stability of the candidate structures and can be used as a
pre-selection tool for choosing the crystal structures to be studied
with the core-blob approach.

The next question is: what happens when crystals of DNACCs melt?  Do
they form a dilute vapour or a dense liquid? To answer this question,
we determined the melting behaviour of the CsCl structure for both
models, and for both we find that the crystal coexists with a dilute
vapour. A similar conclusion was reached in Ref. \citenum{Mar11}.  We
find no evidence for a phase transition between the dilute solution
and a denser liquid phase.  Using the common-tangent construction (see
Fig.~\ref{fig:free_energy}), we can determine the colloidal volume
fraction of the CsCl structure at coexistence, $\eta_{s}$. Within the
core-blob approach, we find $\eta_{s} = 0.029$ at $T=56.9$
$^\circ$C. As expected, the pair potential approach predicts a more
compact equilibrium CsCl crystal of $\eta_{s} = 0.036$. Determining
$\eta_{s}$ within both models for several temperatures, we can then
determine the prediction of the lattice constant $a$ of the
equilibrium CsCl crystals as a function of temperature
\begin{equation}
a(T) = \sqrt[3]{\frac{8 \pi}{3 \eta_{s}(T)}}R_C.
\end{equation}
The computed values of $a(T)$ are compared to experimental data in the
inset of Fig.~\ref{fig:free_energy}. Concentrating first on the
core-blob approach, we find that it predicts the thermal expansion
coefficient of the crystals at least qualitatively correctly. It also
correctly predicts that at sufficiently low temperatures $a$ levels
off to a constant value. However, the simulations predict denser
crystals than experimentally observed.  The discrepancy in lattice
constant is as large as $\sim 12\%$. This observation points to a
problem in the input in our model. One drawback is that we neglected
the hexane-thiol linker grafting the DNA strands to the colloids,
which has an end-to-end length of $\sim 0.8$ nm. Taking this linker
into account is expected to reduce the dispreancy to the experiments
slightly. However, the major weak spots are the choices of the
`preferred' experimental values for the persistence length $p_{\rm
  ss}$ and for the interbase distance $b_0$ of ssDNA. We chose an
average value for both parameters but, in reality, both numbers are
expected to depend on the precise base sequence of the
ssDNA~\cite{Mil99,Kuz01}. With more systematic experimental data on
the sequence dependence of these values, we expect that the core-blob
model would also allow for quantitative predictions of spatial
quantities (such as the lattice constant).

We note that the pair potential approach captures neither the length
nor the temperature behaviour of DNACC crystals correctly: it
seriously overestimates the thermal expansion coefficient of the
crystals.  Hence, the pair-potential approach cannot be used to
describe the thermal properties of DNACC crystals.

For completness, we point out that in the temperature regime
where DNA strands cannot bind anymore (i.e. well above $65$
$^\circ$C), crystallisation can only occur due to excluded volume
effects for which a high osmotic pressure is needed. Since the focus
of our work was on crystal formation triggered by DNA hybridisation,
we did not study this regime.

\begin{figure}%[!b]
\includegraphics[width=8.3cm]{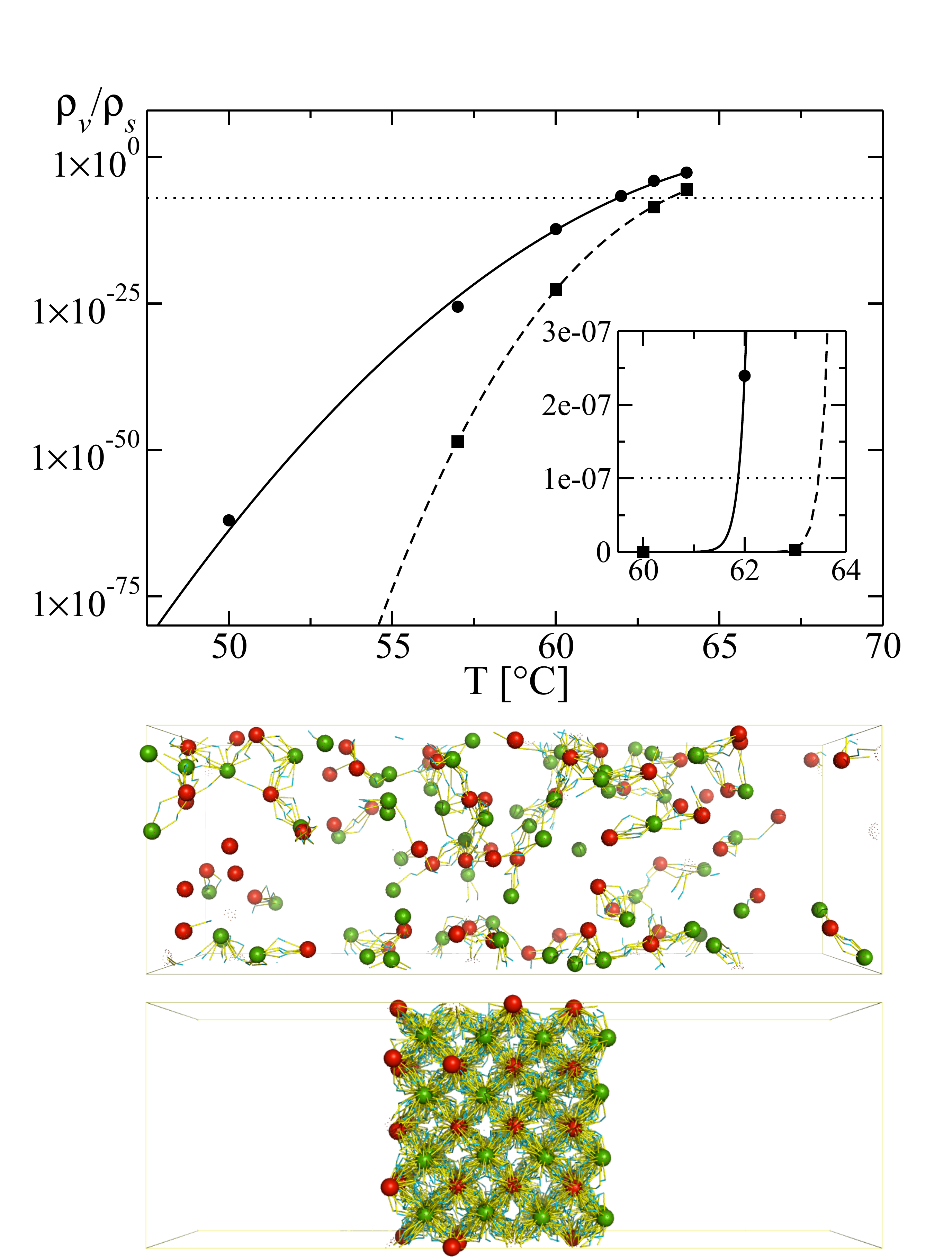}
\caption{Top: The ratio between the coexistence densities of the
  vapour and the solid, $\rho_{v}/\rho_{s}$, as function of the
  temperature according to simulations within the core-blob model
  ($\bullet$) and the pair interaction approach ($\blacksquare$). The
  solid and dashed lines serve as guides to the eye. The threshold
  value of $10^{-7}$ (dotted line) indicates the temperature below
  which spontaneous evaporation of DNACCs from a crystal becomes
  negligible. Middle and bottom: Simulation snapshots of a DNACC
  crystal (core-blob model) at $T = 65.1$ $^\circ$C which melted
  during the course of the simulation and of a crystal at $T=60.2$
  $^\circ$C, where the evaporation of DNACCs is negligible. $A$
  colloids are shown as green spheres, $B$ colloids in red. For
  clarity, only hybridised DNA chains are shown.}
\label{fig:melting}
\end{figure}

The equilibrium melting (`sublimation') temperature of a DNACC crystal
depends on the concentration of the dilute solution and, in contact
with an infinitely dilute solution, all DNACC crystals will eventually
evaporate.  However, the rate at which this happens depends strongly
on temperature. The more relevant question is therefore: at what
temperature does the rate of sublimation of a DNACC crystal become
experimentally observable? Experimentally, the effective melting
temperature was determined via ultraviolet-visible spectrophotometry
in Ref. \citenum{Nyk08} and found to be $T_m = 62.5 (\pm 0.3)$
$^\circ$C. To be able to compare simulational data to experiments, we
can estimate the temperature below which the spontaneous evaporation
of DNACCs from a crystal becomes negligible. A rough estimate of the
concentration where this happens can be obtained using Smoluchowski's
treatment of the diffusion limited growth of a
cluster~\cite{Smo17,Bam85}. In equilibrium, the evaporation rate of
DNACCs from a solid, spherical cluster of radius $\hat R$ equals the
aggregation rate to that cluster in the presence of a dilute vapour of
DNACCs of density $\rho_{v}$. We can determine the latter rate, ${{\rm
    d}N}/{{\rm d}t}$, as
\begin{equation}
\frac{{\rm d}N}{{\rm d}t} = 4\pi\rho_{v} \hat D \hat R,
\end{equation}
with $\hat D$ being the diffusion constant of a DNACC
\cite{Smo17,Bam85}. On the other hand, the amount of DNACCs in the
solid cluster can be expressed as $N=\frac{4\pi}{3} \hat R^3\rho_{s}$, where $\rho_{s}$ is the density of the
cluster and $\rho_s = \eta_s/(\frac{4\pi}{3} R_C^3)$. Then, the following
relation between $\rho_{v}$ and $\rho_{s}$ can be derived:
\begin{equation}
\frac{\rho_v}{\rho_s} = \frac{1}{2\hat D} \frac{{\rm d}\hat R^2}{{\rm
    d}t}
\end{equation}
To estimate $\rho_v$, we assume that a solid cluster grows to 1 $\mu$m
in one day ($\sim 10^5$s). Further, we estimate the diffusion constant
of a DNACC from the Einstein--Smoluchowski relation, $\hat D =
k_BT/(6\pi\hat\eta R_C)$. With $\hat \eta \sim 1$cP being the
viscosity of water, we find that $\hat D\sim 4\times
10^{-7}$cm$^2$/s. Then, $\rho_{v}/\rho_{s} \sim 10^{-7}$. From the
knowledge of the coexistence densities of the vapour and solid phases
at several temperatures, we find that the last relation is fulfilled
for $T_m=63.5 (\pm 0.2)$ $^\circ$C in the pair potential approach and
for $T_m = 61.9 (\pm 0.2)$ $^\circ$C in the core-blob model (see
Fig. \ref{fig:melting}), which is in good agreement with the
experimental finding of $T_m=62.5 (\pm 0.3)$ $^\circ$C. We note that
in Ref. \citenum{Mla12}, we estimated the melting temperature within
the core-blob approach by considering the point where small colloidal
crystals melted on the time scale of a simulation. That approach is
likely to lead to a higher estimate of the melting temperature and,
indeed, we found that simulated crystals melt for $T \gtrsim 64.3 (\pm
0.5)$ $^\circ$C (see Fig.~\ref{fig:melting}).

\section{Conclusion}\label{sec:con}
In this paper we described a method to construct a quantitative
coarse-grained model of DNACCs and compared its results to
experimental data. For comparison, we also studied the predicitve
power of a simpler model based on the effective interactions between
pairs of DNACCs.

We found that the pair potential approach can be used as a qualitative
tool, allowing to roughly delimit the range of mechanical stability of
DNACCs, and give a coarse estimate of the melting
temperature. Further, it allows for qualitative insight into the
compression behaviour of DNACC crystals upon temperature
reduction. This model has the advantage of being computationally
inexpensive, therefore allowing for fast and large-scale testing of
DNACC designs.

However, we showed that for quantitative insight into DNACC systems in
the regime where the radius of gyration of the tethered DNA strands is
of the order of the size of the (nano-)colloid, an explicit
description of sticky ends is needed to capture the competition of
DNACCs for DNA bridges. We therefore developed a more detailed model,
which we termed ``core-blob model'' and which is solely based on
experimental input. Results from this approach show good agreement
with experimental data in temperature-dependent quantities. While the
lattice constant is not captured quantitatively in absolute values,
the thermal expansion coefficient of crystals is well described.  We
speculate that, once more systematic experimental data on the
persistence length and inter-base distance of ssDNA become available,
the core-blob model should also account for the experimentally
observed lattice spacings. We could of course have adjusted the model
parameters to account for the observed lattice spacing, but this would
have defeated the purpose of the present work, which was to construct
a model based exclusively available ssDNA data - our model contains no
{\it a posteriori} fitted parameters.

A potential drawback of the core-blob model is that the input
parameters depend on the ssDNA length and sequence.  For a given set
of input parameters, new simulations are needed to redevelop the
interaction potentials of the core-blob model and to predict the phase
behaviour.  In some cases, the situation may be better. For instance,
in e.g.~Ref. \citenum{Xio09}, different systems were generated by
supplementing {\it one} choice of DNACCs with different linkers. Then,
the length of spacers, their binding strength and the number of
reactive ends can be tuned in a straightforward way via these linker
sequences alone. Similarly, our model can be easily generalised to
incorporate linkers while reusing the present representation of the
DNACCs themselves. Furthermore, the core-blob model can be adapted to
study e.g.~systems of more complex coatings, asymmetric mixtures, or
polydisperse systems, while allowing for direct mapping to the
corresponding experimental system.

In summary, a rough scanning of the phase behaviour of DNACC designs
via the pair potential approach can be used to preselect promising
DNACC designs, which can then be quantitatively studied in more
detailed, but also computationally more expensive calculations using
the core-blob model. In this way, the approach presented here offers a
path to computer-aided design of suitable DNA-grafted building blocks,
advancing the efforts of constructing truly complex self-assembling
structures.

\section{Acknowledgements}
We thank P.~Varilly for helpful discussions and careful reading of the
manuscript. Further, we thank B.~Capone (Vienna),
S.~Angioletti-Uberti, B.~M.~Mognetti, W.~Jacobs, G.~Day (Cambridge),
A.~Tkachenko, D.~Nykypanchuk and O.~Gang (Brookhaven) for useful
discussions at various stages of this project. BMM acknowledges EU
funding (FP7-PEOPLE-IEF-2008 No.~236663 and FP7-PEOPLE-CIG-2011
No.~303860) and funding via the MFPL Vienna International
Post-Doctoral Program for Molecular Life Sciences (funded by Austrian
Ministry of Science and Research and City of Vienna, Cultural
Departement - Science and Research).  AD was supported by an EMBO
longterm fellowship. DF and FJMV acknowledge support of ERC Advanced
Grant 227758. DF acknowledges a Wolfson Merit Award of the Royal
Society of London and EPSRC Programme Grant EP/I001352/1.

\section{Appendix A: The binding move in the core-blob model}
In the following, we derive the Monte Carlo algorithm of the binding
move used for hybridisation of sticky ends in the core-blob model. In
Sec.~\ref{app_a}, we study the requirement of detailed balance for the
case of an unbound sticky end binding to a complementary, unbound
sticky end within reach, i.e.~within distance of hybridised sticky
ends, ${\mathcal{L}}$. Without loss of generality, we name the chosen
sticky end $a$ and assume that it is attached to a colloid of kind
$A$. Consequently, we term its possible binding partner $b$, which is
fixed on a colloid of kind $B$. We then generalise to an arbitrary
number of possible binding partners (incl.~the possibility of a
partner change) in Sec.~\ref{app_b}.

\subsection{Detailed balance for one possible binding
  partner}\label{app_a} To fulfil detailed balance, we need to justify
that the flow $\mathcal{K}$ from the configuration where $a$ is
unbound (``$a$'') to the state where $a$ and $b$ are hybridised
(``$ab$'') is the same as the reverse flow, i.e.
\begin{equation}\label{eqn:fluxes}
  \mathcal{K}(a \rightarrow ab) = \mathcal{K}(ab \rightarrow a).
\end{equation}
We can write
\begin{equation}\label{eqn:Kab}
\mathcal{K}(a \rightarrow ab) = P_a P_{\rm gen}(a \rightarrow ab) P_{\rm
  acc}(a \rightarrow ab),
\end{equation}
where $P_a$ is the probability of $a$ being unbound, $P_{\rm
  gen}(a \rightarrow ab)$ is the probability that the Monte Carlo move
 hybridises $a$ with $b$ and $P_{\rm acc}(a
\rightarrow ab)$ is the probability of accepting this move.
An analogous formula can be written for $\mathcal{K}(ab \rightarrow a)$. 

Next, we write the various terms in Eqn.~\ref{eqn:Kab} in terms of the
coordinates of the chosen sticky end $a$ that we try to bind, ${\bf
  r}_a$, and the distance of $a$ from its possible binding partner
$b$, ${r}_{a b} = |{\bf r}_b - {\bf r}_a|$.  All other coordinates
will be denoted by $\{{\bf r}_{\rm rest}\}$.

The probability to be in the unbound state is then given by
\begin{equation}
P_a = e^{-\beta U_{a}} d{{\bf r}_a} r_{a b}^2  
 d{r}_{a b} d\Omega \{d{\bf r}_{\rm rest}\} \frac{q_a^{\rm
    int}q_b^{\rm int}}{\Lambda_a^3 \Lambda_b^3}
\end{equation}
where $ r_{a b}^2 d{r}_{a b} d\Omega$ is an infinitesimal volume
element around the location of $b$. The potential energy of the
state where $a$ is unbound is given by $U_a$, while $q_a^{\rm int}$ and
$q_b^{\rm int}$ are the internal partition functions of $a$ and $b$.
$\Lambda_a$ and $\Lambda_b$ denote the respective de Broglie
wavelengths.

When the sticky ends $a$ and $b$ hybridise, we place ${\bf r}_{b}$
along the connection line ${\bf r}_{a b}$ at a distance $\mathcal{L}$
from ${\bf r}_a$. The coordinate of $a$ remains unchanged. Since there
is only one way to implement this move, the generation probability is
$P_{\rm gen}(a \rightarrow ab) = 1$.

The probability to be in the hybridised state, $P_{ab}$, where $a$ and
$b$ are connected by a rod of length $\mathcal{L}$, is given by
\begin{equation}
P_{ab} = e^{-\beta U_{ab}} d{{\bf r}_a} d\Omega \{d{\bf r}_{\rm
  rest}\} \frac{q_{ab}^{\rm int, rest}}{\Lambda_{ab}^3}.
\end{equation}
where $U_{ab}$ is the potential energy of the state where $a$ and $b$
are hybridised. The internal partition function of the hybridised
sticky ends which are restrained in their rotational freedom is
denoted by $q_{ab}^{\rm int, rest}$. Since $q_{ab}^{\rm int, rest}$ is
independent of the precise orientation of the hybridised sticky ends in space,
it can be related to the internal partition of a rotationally
unrestricted dsDNA segment, $q_{ab}^{\rm int}$, via $q_{ab}^{\rm int,
  rest} = q_{ab}^{\rm int}/4 \pi$ and thus
\begin{equation}
P_{ab} = e^{-\beta U_{ab}} d{{\bf r}_a} d\Omega \{d{\bf r}_{\rm
  rest}\} \frac{q_{ab}^{\rm int}}{4 \pi \Lambda_{ab}^3}.
\end{equation}

The factor $4 \pi$ results from the fact that the connection rod
between $a$ and $b$ is restrained from its rotational
freedom. Further, $\Lambda_{ab}$ is the de Broglie wavelength.

To generate the unbound state from the hybridised one, we have to
generate a new coordinate for $b$ along the connection line of the
bound $b$ and $a$, ${\bf r}_{ab}$. Generating the new position with a
probability proportional to the distance squared, we get
\begin{equation}
P_{\rm gen}(ab \rightarrow a) = \frac{r_{ab}^2
  d{r}_{ab}}{\frac{1}{3}\mathcal{L}^3}
\end{equation}
By imposing Eqn.~\ref{eqn:fluxes}, we get the condition for the
acceptance probabilities for the Monte Carlo move:
\begin{equation}
\frac{P_{\rm acc}(a \rightarrow ab)}{P_{\rm acc}(ab \rightarrow a)} =
\frac{K}{\frac{4 \pi}{3}\mathcal{L}^3 \rho_0} e^{-\beta \Delta U},
\end{equation}
with $\Delta U = U_{ab}-U_a$. In this last equation, we have used that
\cite{BozPhD}
\begin{equation}
\frac{q_{ab}^{\rm int} \Lambda_a^3 \Lambda_b^3}{q_a^{\rm int} q_b^{\rm
    int} \Lambda_{ab}^3} = \frac{K}{\rho_0},
\end{equation}
with $K$ being the equilibrium binding constant and $\rho_0$ the
standard density of 1 mol/l.

\subsection{Algorithm for the binding move in the case of several possible binding partners}\label{app_b}
\begin{figure}%[!b]
\includegraphics[width=8.3cm]{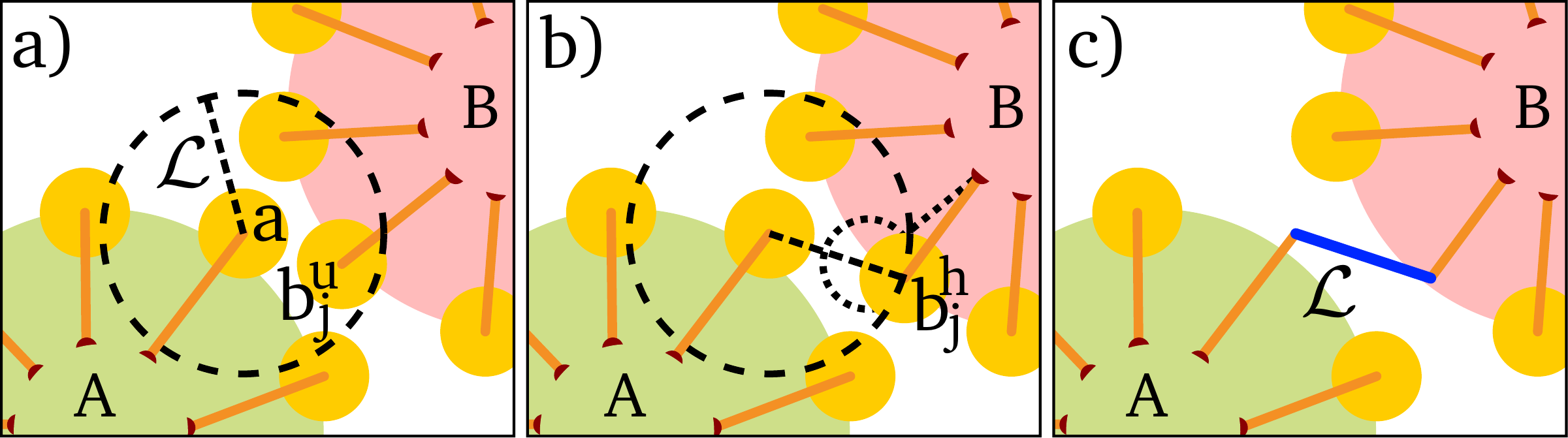}
\caption{(Colour online) The hybridisation move. a) From a chosen blob
  $a$ on colloid $A$, all possible (unbound) binding partners $b_j$
  are determined which are within a sphere of radius of the hybridised
  sticky ends, ${\mathcal{L}}$. b) A binding partner, $b_j$, is chosen
  according to its weight (see text). This sticky end is moved from
  its unbound coordinates, $b_{j}^u$, to its coordinates in the
  hybridised state, $b_j^h$. c) The DNA bridge formed by the
  hybridised sticky ends is modelled as a rigid rod of fixed length
  (blue rod). Note that the figure disregards the case where $a$ is
  originally bound.}
\label{fig:binding_move}

\end{figure}

Here, we outline the algorithm for the Monte Carlo binding/partner
change move for several possible binding partners.
\begin{enumerate}
\item Choose a blob (bound or unbound) at random. Without loss of
  generality, we assume that this blob is on a colloid of kind $A$.
  Therefore, we denote this blob as $a$ and its coordinates as ${\bf
    r}_{a}$.

\item[2a.] Find all $j = 1,\dots, N_{p}^u$ {\it unbound} blobs on
  unlike colloids (here: kind $B$), $\{b_j\}$, that are within
  distance $\mathcal{L}$, i.e.~$|{\bf r}_{ab_j^u}| = |{\bf
    r}_{b_j^u}-{\bf r}_a| \leq \mathcal{L}$ (see
  Fig.~\ref{fig:binding_move}a). Here, ${\bf r}_{b_j^u}$ is the
  coordinate of the unhybridised possible binding partner $j$.

\item[2b.] In case $a$ is initially bound, add its actual binding
  partner to this list; the coordinate of this binding partner is
  denoted as ${\bf r}_{b_j^h}$ since it is hybridised with $a$. We
  then have a total of $N_p^u+1$ possible binding partners.

\item[3a.] For all $N_p^u$ {\it unbound} binding partners $b_j$ at
  their unbound positions ${\bf r}_{b_j^{u}}$, do the following: along
  the connection line ${\bf r}_{ab_j^u}$, generate the new position
  ${\bf r}_{{b_j^h}}$ as if $b_j$ were hybridised with $a$ (see
  Fig.~\ref{fig:binding_move}b), i.e.~place it at distance
  $\mathcal{L}$ from $a$:
\begin{equation}
{\bf r}_{b_j^{h}} = {\bf r}_{a} + \mathcal{L} {\bf
  r}_{ab_j^{u}} / |{\bf r}_{ab_j^{u}}|.
\end{equation}

\item[3b.] If $a$ was hybridised initially, further compute the
  following for its actual binding partner $b_j$: Along the connection
  line ${\bf r}_{ab_j^h}$, randomly generate a new, unbound position
  of $b_j$, ${\bf r}_{b_j^u}$ as
\begin{equation}
{\bf r}_{b_j^{u}} = {\bf r}_{a} + \sqrt[3]{x} \left({\bf
  r}_{b_j^{h}}-{\bf r}_{a}\right).
\end{equation}
with random number $x \in [0,1)$.
\end{enumerate}
After this step, both ${\bf r}_{ab_j^h}$ and ${\bf r}_{ab_j^u}$ are
known for all possible binding partners.

\begin{enumerate}
\item[4a.] Calculate the weight of the state where $a$ is not
  hybridised as
\begin{equation}
W_a = \exp(-\beta U_a),
\end{equation}
where
\begin{equation}
U_a = \sum_j U_{b_j^{u}}.
\end{equation}
In the last equation, $U_{b_j^{u}}$ is the energy that each of the $j$
possible binding partners at their unbound positions have with the
rest of the system, including the repulsion of $a$. Care has to be
taken not to double count interactions between the various $b_j$.

\item[4b.] Calculate the weights for each of the possible hybridised
  states. The weight of the state where $b_j$ is bound to $a$ is given
  as
\begin{equation}
W_{ab_j} = \frac{K}{\frac{4 \pi}{3} {\mathcal L}^3 \rho_0} \exp(-\beta
U_{ab_j}),
\end{equation}
where
\begin{equation}
U_{ab_j} = U_{ab_j^{h}} + \sum_{j's(\neq j)} U_{b_{j'}^{u}},
\end{equation}
where $U_{b_{j'}^{u}}$ are the energies that each of the $j'(\neq j)$
possible binding partners at their unbound positions have with the
rest of the system, including the repulsion of $a$ and $b_j$ at their
bound positions. Further, $U_{ab_j^{h}}$ is the interaction of $b_j$
at its bound position with the rest of the system, including
$a$. Again, care has to be taken not to double count interactions.

\item[5.] Calculate the sum $S$ over all weights
\begin{equation}
S = W_a + \sum_j W_{ab_j} .
\end{equation}

\item[6.] Randomly choose a state according to the probabilities of
  the unbound state
\begin{equation}
P_a = \frac{W_a}{S},
\end{equation}
and the various hybridised states
\begin{equation}
P_{ab_j} = \frac{W_{ab_j}}{S}.
\end{equation}
If a hybridised state $ab_j$ is chosen, $a$ and $b_j$ are connected by
a rod and cannot move independently anymore (see
Fig.~\ref{fig:binding_move}c).

\end{enumerate}
\section{Appendix B: Fitting functions for effective interactions}
To use effective interactions in computer simulations, it is
crucial to find reliable fits for both $\Phi_{\rm 2,rep}$ and
$\Phi_{\rm 2,hyb}$. The former can be fitted by a
sigmoidal curve for large distances $r>5.5R_C$
\begin{equation}
\beta \Phi_{\rm 2,rep}^{\rm fit}(r>5.5R_C) =
\frac{a_1}{1+\exp\left(-\frac{r-b_1}{c_1}\right)},
\end{equation}
with fitting parameters $a_1$, $b_1$ and $c_1$. For distances
$r<5.5R_C$, the potential can be approximated by an exponential
function
\begin{equation}
\beta \Phi_{\rm 2,rep}^{\rm fit}(r<5.5R_C) = a_2
\exp\left(-\frac{r}{b_2}\right)+c_2
\end{equation}
with fitting parameters $a_2$, $b_2$ and $c_2$. The fits have to be
performed under the boundary condition that the two functions need to
join smoothly at $r=5.5 R_C$.

The attractive potential $\Phi_{\rm 2,hyb}$ is fitted in two steps:
for distances where DNA strands can hybridise, i.e.~$r \lesssim 6
R_C$, the potential is approximately linear, $\Phi_{\rm 2,hyb} \sim
kx+d$. For distances $r \gtrsim 7.5 R_C$, $\Phi_{\rm 2,hyb} =
0$. Then, $\Phi_{\rm 2,hyb}$ can be fit for each temperature of
interest as an interpolation between these two trends \cite{Ker10},
i.e.
\begin{equation}
\beta \Phi_{\rm 2,hyb}^{\rm fit} (r; T~{\rm fixed}) = kx + \tau k \ln
\left[1+\exp\left(\frac{\alpha-x}{\tau} \right) \right] + d
\end{equation}
with $\alpha = -d/k$ and fitting parameter $\tau$.

\bibliographystyle{rsc} \bibliography{cambridge}

\end{document}